\documentclass[reprint,
superscriptaddress,
nofootinbib,
aps,
pra,
showkeys
]{revtex4-2}
\pdfoutput=1

\usepackage{amsthm, mathtools, amssymb, mathrsfs, dsfont}
\usepackage{braket, physics} 
\usepackage{siunitx}
\usepackage{orcidlink}

\usepackage{soul} 
\usepackage{algorithm}
\usepackage{algorithmicx}
\usepackage[noend]{algpseudocode}

\usepackage{graphicx, placeins, float}
\usepackage[caption=false]{subfig}
\usepackage[export]{adjustbox}
\graphicspath{ {./Figures/} }

\usepackage{hyperref,xcolor}
\usepackage[nameinlink,capitalize]{cleveref}
\hypersetup{
	colorlinks = true,
	linkbordercolor = {white},
	linkcolor={purple},
	citecolor={purple},
	urlcolor={blue}
}

\makeatletter
\renewcommand{\ALG@name}{Algorithm}
\makeatother
\crefname{algorithm}{Alg.}{Algs.}
\newcommand{\projector}[3]{\ket{#1}_{#3}\!\bra{#2}}

\newcommand{\bigO}{\mathcal{O}}

\begin{document}
\title{Quantum frequency resampling}
  \author{Emanuele Tumbiolo\,\orcidlink{0009-0000-6411-1285}}
	\email[Emanuele Tumbiolo: ]{emanuele.tumbiolo01@ateneopv.it}
    \affiliation{Dipartimento di Fisica, Università degli Studi di Pavia, Via Agostino Bassi 6, I-27100, Pavia, Italy}
    \affiliation{INFN Sezione di Pavia, Via Agostino Bassi 6, I-27100, Pavia, Italy}
	
\author{Simone Roncallo\,\orcidlink{0000-0003-3506-9027}}
	\email[Simone Roncallo: ]{simone.roncallo01@ateneopv.it}
    \affiliation{Dipartimento di Fisica, Università degli Studi di Pavia, Via Agostino Bassi 6, I-27100, Pavia, Italy}
    \affiliation{INFN Sezione di Pavia, Via Agostino Bassi 6, I-27100, Pavia, Italy}
    
\author{Chiara Macchiavello\,\orcidlink{0000-0002-2955-8759}}
	\email[Chiara Macchiavello: ]{chiara.macchiavello@unipv.it}
    \affiliation{Dipartimento di Fisica, Università degli Studi di Pavia, Via Agostino Bassi 6, I-27100, Pavia, Italy}
    \affiliation{INFN Sezione di Pavia, Via Agostino Bassi 6, I-27100, Pavia, Italy}

\author{Lorenzo Maccone\,\orcidlink{0000-0002-6729-5312}}
	\email[Lorenzo Maccone: ]{lorenzo.maccone@unipv.it}
    \affiliation{Dipartimento di Fisica, Università degli Studi di Pavia, Via Agostino Bassi 6, I-27100, Pavia, Italy}
    \affiliation{INFN Sezione di Pavia, Via Agostino Bassi 6, I-27100, Pavia, Italy}

\begin{abstract}
   In signal processing, resampling algorithms can modify the number of resources encoding a collection of data points. Downsampling reduces the cost of storage and communication, while upsampling interpolates new data from limited one, e.g. when resizing a digital image. We present a toolset of quantum algorithms to resample data encoded in the probabilities of a quantum register, using the quantum Fourier transform to adjust the number of high-frequency encoding qubits. We discuss advantage over classical resampling algorithms.
\end{abstract}
\maketitle

\section{Introduction\label{sec:I}}
Signal resampling (or sample-rate conversion) can modify the rate at which continuous processes are discretized into a finite set of data points \cite{book:Oppenheim2011discrete}. It consists of two reciprocal operations: downsampling, which simplifies the signal representation by reducing such rate, and upsampling, which interpolates existing data to a higher-dimensional representation. Classically, these techniques are typically used to improve the compatibility between systems with different data formats. Several protocols perform resampling in the frequency domain by three steps: (i) signal conversion, e.g. using the fast Fourier transform (FFT), (ii) manipulation of the high-frequency components, and (iii) conversion back to the original domain, e.g. using the inverse FFT.

Quantum effects can be employed to tackle signal processing tasks, efficiently manipulating both quantum data and classical information.
In practice, data can be constrained by hardware or algorithmic requirements. Communication through a limited-capacity channel requires downsampling into fewer qubits, while a variational circuit (trained on a different qubit count) must resample inputs with different size. Eventually, one could measure, resample and re‐prepare the desidered state, with a cost that scales at least linearly in the number of samples. Such operations — plus state‐preparation overhead — are resource-hungry for large datasets.
In quantum image processing
\cite{art:Venegas,art:Yan,art:Zhang_NEQR,art:Le,book:Yan},  several compression and interpolation algorithms have been proposed, e.g. based on matrix product state truncation \cite{art:Latorre}, bilinear interpolation \cite{art:Zhou,art:Yan}, quantum variational circuits \cite{art:Peng} and Fourier methods \cite{art:RoncalloQJPEG}.
Resampling algorithms have also been investigated in quantum state preparation \cite{art:Kitaev_GaussianWavefuncs,art:RamosQuInterpolation,art:Stefanski}. However, a general framework for quantum signal resampling is still lacking.
\begin{figure*}
	\centering
	\includegraphics[width=0.7\textwidth]{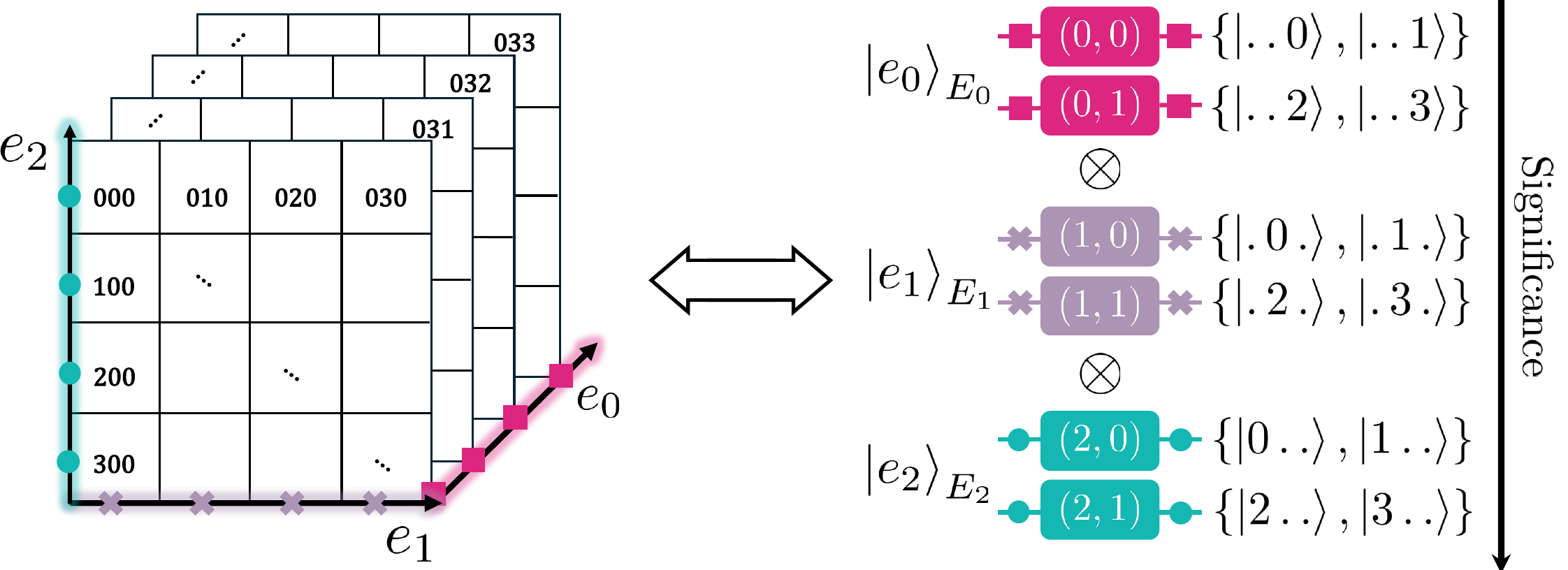}
\caption{\label{fig:AmplitudeEnc}Little-endian ordering when encoding a 3D signal $\mathcal{S}$. Classically, $\mathcal{S}$ is represented by a $4\times4\times4$ array, with a total of $64$ samples arranged along axes $e_{0}, e_{1}, e_{2}$. According to \cref{eq:Input}, these values are encoded in the probabilities of a $6$ qubits register $E$, which can be decomposed into $3$ subregisters, $E_0$, $E_1$, $E_2$, arranged, top to bottom, in order of increasing significance.}
\end{figure*}

In this paper, we address this gap by introducing two protocols (agnostic to signal type and dimensionality) that tackle the fundamental resampling operations - downsampling and upsampling - for data encoded in the probabilities of a multiqubit state \cite{book:Petruccione}. These techniques establish a framework for multidimensional resampling in the frequency domain, of which \cite{art:RoncalloQJPEG} is an application for the downscaling of digital images. Using a combination of Hadamard gates and quantum Fourier transforms (QFTs) \cite{book:Nielsen,art:Hales}, we achieve downsampling by discarding the most significant qubits of the register (corresponding to the fastest-varying components of the signal): this yields a lossy compression, that allows a faithful interpolation when such frequencies are negligible. Conversely, upsampling appends new qubits to increase the signal resolution via redundancy: this process is invertible and preserves the original information content. We analyze such protocols from an analytical and numerical point of view, deriving closed-form expressions for their outputs. Detailed calculations are reported in the Appendix. This is a two-fold result, that identifies the classical equivalents of our algorithms as well as their advantages and limitations. We discuss the resource cost of these methods, both in terms of gates and experimental repetitions. For gates, algorithms achieve an exponential speedup over their classical counterparts. Moreover, we establish the robustness of downsampling against the statistical cost of output retrieval.

\section{The algorithms \label{sec:II}}
In this section we discuss the operating principles of our quantum resampling algorithms for a signal encoded in the probabilities of a multiqubit state. We propose two complementary protocols, one for downsampling, i.e. for reducing the number of samples of the signal while preserving its pattern, and one for upsampling, namely for interpolating new information from the existing one. Both algorithms employ a multidimensional quantum Fourier transform (MD-QFT) to process the input in the frequency domain, where we either discard the qubits encoding the highest frequency modes of the signal or add fresh qubits to the register, expanding the signal spectrum with vanishing high-frequency components. Here, frequency describes how rapidly data features oscillate, with respect to the Fourier-basis representation of the signal. Low frequency modes capture smooth, slowly varying structures, while
high frequencies correspond to sharp transitions.

A discrete signal $\mathcal{S}$ is a multidimensional array of $d$ indexes (i.e. coordinates, called axes), namely a collection of finite values, sampled from an underlying continuous process at a given sampling rate and distributed in a $d$-dimensional grid.  For instance, a digital image can be described as a two-dimensional signal, where the samples correspond to the pixel intensities, and the axes define a coordinate system on the image plane. Similarly, a video can be viewed as a three-dimensional array, with two spatial axes and a temporal one that orders the sequence of frames in time. Each axis may have a distinct sampling rate, specifying the average number of samples acquired in the unit interval \cite{book:MoirRudiments}.   We limit our discussion to hyper-cubic, i.e. square, arrays, with $N_0$ entries per axis and thus a total of $N_0^d$ samples, since the extension to different entries per axis is straightforward. We employ an amplitude encoding scheme \cite{book:Petruccione} to upload $\mathcal{S}$ onto the state of an \textit{encoding register} $E$ of $n_E = \log_2(N_0^d)$ qubits, as
\begin{align}
    \label{eq:Input}
    \ket{\Psi}_E &= \sum_{\mathbf{e}}\alpha_{\mathbf{e}} \ket{\mathbf{e}}_E\,\, ,
    \intertext{where $\ket{\mathbf{e}}_E$ labels the computational basis elements and}
    \alpha_{\mathbf{e}}  &= \sqrt{\frac{\mathcal{S}_{\mathbf{e}}}{I_E}}\;\;,
\end{align}
with $S_\mathbf{e}$ the sample values normalized to the total input signal intensity $I_E = \sum_{\mathbf{e}}\mathcal{S}_\mathbf{e}$. Unless otherwise specified, summations are always taken over all the possible values of their index.
The $d$-tuple $\mathbf{e} = (e_0,e_1,...,e_{d-1})$, with $e_i \in \{0,N_0 - 1\}$, labels the entries of $\mathcal{S}$ along each dimension. Due to the correspondence between sample indexes and computational basis elements, $E$ can be decomposed in $d$ subregisters $E_i$ of $n_0$ qubits, one for each signal axis, such that $n_E = d\, n_0$. In the computational basis 
\begin{equation}
    \ket{\mathbf{e}}_E = \ket{e_{d-1}}_{E_{d-1}}\otimes ... \otimes \ket{e_0}_{E_0} \,\, .
    \label{eq:s_basis}
\end{equation}
We call $\ket{\Psi}_E$ \textit{quantum signal}. Qubits in $E$ are arranged from top to bottom in order of increasing significance, following the so-called little-endian ordering \cite{book:Patterson_LittleEndian}. Hence, the $E_{d-1}$ subregister contains the $n_0$ most significant qubits, i.e. the entries along the first dimension and so on, as depicted in \cref{fig:AmplitudeEnc}. 
We identify each qubit with a pair of indices $(s,q)$, where $0\leq s \leq d-1$ labels the subregister, and $0\leq q\leq n_0-1$ the position within it. 
\begin{figure*}
    \subfloat[]{\includegraphics[valign = c, width = 0.55 \textwidth]{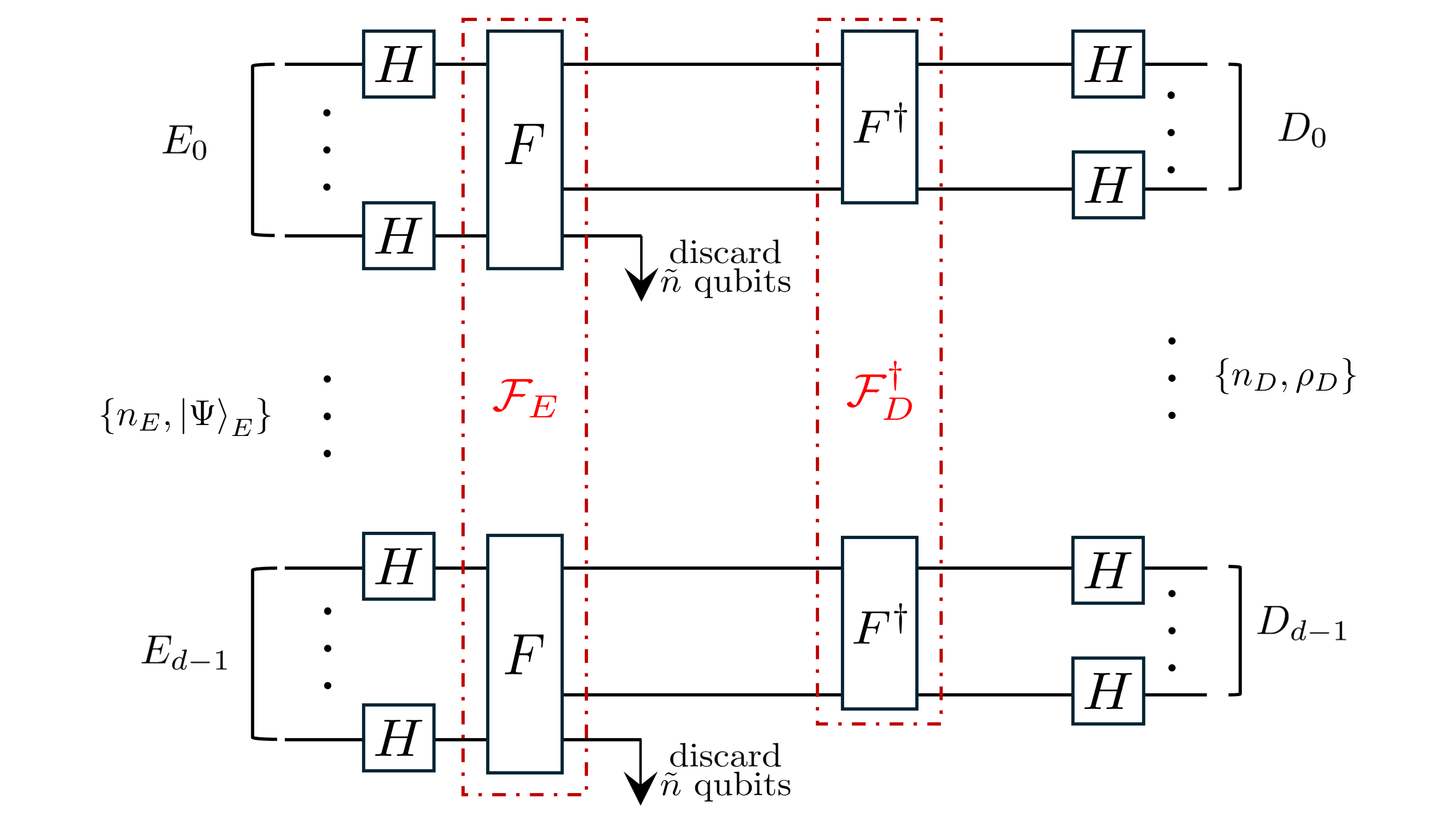} \label{fig:DownCircuit}}
    \subfloat[]{\includegraphics[valign = c, width = 0.425 \textwidth]{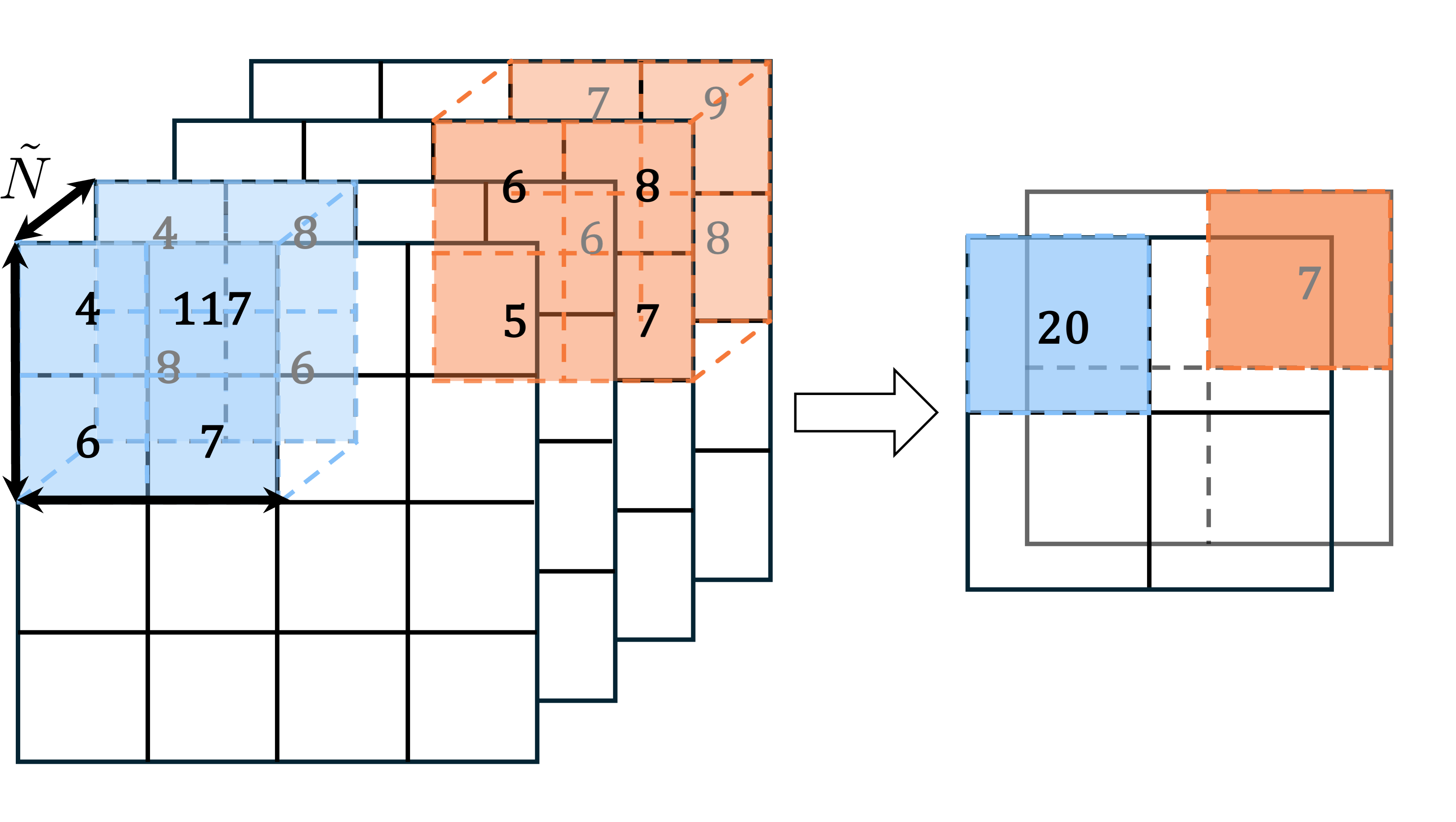} \label{fig:BlockAverage}}\caption{\label{fig:DownAlgorithm_Rapr} Quantum downsampling of a $d$-dimensional signal encoded in the probabilities of a register $E$, divided in $d$ subregisters $E_i$, of $n_0$ qubits each. (a) Circuit scheme of the algorithm, which discards $\tilde{n} < n_0$ qubit for each axis, compressing the input into a state of register $D$, with $n_D = n_E - d\,\tilde{n}$ qubits (consistently divided into $d$ subregisters). (b) Effect on the classical signal. The discarding parameter $\tilde{n}$ determines the output resolution, from $N_0^d$ to $N_0^d/2^{d\tilde{n}}$.The algorithm is equivalent to averaging the original signal on a set of hyper-cubic blocks of side $\Tilde{N}$.}
\end{figure*}

In this representation, we define the MD-QFT $\mathcal{F}$ as the tensor product of $d$ standard, i.e. one-dimensional, quantum Fourier transforms $F$ \cite{book:Nielsen}, each acting on a separate subregister. For example, on the encoding register
\begin{equation}
    \begin{split}
     \mathcal{F}_E\ket{\mathbf{e}}_E &= \,F\ket{e_{d-1}}_{E_{d-1}}\otimes...\otimes F\ket{e_{0}}_{E_{0}} =\\
    &=N_0^{-d/2}\sum_{\mathbf{k}} \exp{2\pi i \mathbf{k}\cdot \mathbf{e}/N_0} \ket{\mathbf{k}}_E \,\, ,
\end{split}
\end{equation} with $\mathbf{k} = (k_0,k_1,...,k_{d-1})$, $k_i \in \{0,...,N_0-1\}$, $\mathbf{k}\cdot \mathbf{e}$ the usual dot product and $i$ the imaginary unit.
Similarly, its inverse is \begin{equation}
    \begin{split}
    \mathcal{F}_{E}^{\dagger} \ket{\mathbf{e}}_E &= F^{\dagger}\ket{e_{d-1}}_{E_{d-1}} \otimes ... \otimes F^{\dagger} \ket{e_{0}}_{E_{0}} = \\
     &= N_0^{-d/2} \sum_{\mathbf{k}} \exp{-2\pi i \mathbf{k}\cdot \mathbf{e}/N_0}\ket{\mathbf{k}}_E\,\, .
    \end{split}
\end{equation}

\textbf{Quantum downsampling}
 We now focus on the downsampling algorithm, which reduces the size of the encoding register from $n_E$ to $n_{D} = d \,n_1$, with $n_1 = n_0 - \tilde{n}$ the number of qubits per axis in the \textit{downsampled register D}. The settable parameter $\tilde{n}$ controls the downsampling ratio
\begin{equation}
    \label{eq:CompressionRatio}
    \frac{n_E}{n_D} = 1 + \frac{d\tilde{n}}{n_D}\,\,
\end{equation}of the output signal. In the following, we discuss how this is related to output quality. 

The algorithm works as follows. Apply a set of Hadamard gates $H^{\otimes n_E}$ to $E$ - namely an Hadamard gate $H$ to each encoding qubit -  followed by a MD-QFT, which converts the state of the register into the Fourier domain. For each of the $d$ subregisters, discard (by partial tracing) the $\Tilde{n}$ most significant qubits - i.e. those corresponding to the highest powers of two in the binary expansion of the basis index -  retaining only those with $0\leq q \leq (n_0-\Tilde{n}) -1$ in the downsampled register. This operation amounts to averaging out the high-frequency components of the signal and yields a reduced state of $n_{D}$ qubits. Consequently, the number of samples in the signal is uniformly reduced  along each axis by a factor of $\widetilde{N}=2^{\tilde{n}}$. 
Take the inverse MD-QFT $\mathcal{F}_{D}^{\dagger}$ on the reduced register.
Finally, apply a set of Hadamard gates $H^{\otimes n_D}$ to the remaining qubits, thus returning to the computational basis. We report the circuital implementation of \cref{alg:Downsampling} in \cref{fig:DownCircuit}. The output, now encoded by the density operator $\rho_D$, has only $N_1 = N_0/\widetilde{N}$ samples per axis, and thus a total of $N_1^d$ entries. In terms of sampling rates, this is equivalent to introducing a reduced effective rate  $\mathsf{f}_i^D = \mathsf{f}^E_i/\Tilde{N}$, where $\mathsf{f}_i^E$ is the input rate for the $i$-th axis. Such samples can be recovered from the probability distribution $p_{\mathbf{m}} = \text{Tr}[\rho_D \projector{\mathbf{m}}{\mathbf{m}}{D}]$ where $\ket{\mathbf{m}}_D = \ket{m_{d-1}}_{D_{d-1}}\otimes...\otimes \ket{m_0}_{D_0}$ is the computational basis element of the downsampled register.
These probabilities are related to the input by
\begin{equation}
    \label{eq:DownOut}
    p_{\mathbf{m}} = \sum_{\mathbf{\Tilde{e}}}\frac{ \mathcal{S}_{\Tilde{N}\mathbf{m}+\Tilde{\mathbf{e}}}}{I_D} \,\, , \end{equation}
where $I_D = \tilde{N}^{-d}/I_E$ is the downsampled signal intensity, $\mathbf{\Tilde{e}} = (\Tilde{e}_0,\dots,\Tilde{e}_{d-1})$, $\Tilde{e}_i \in \{0,\Tilde{N}-1\}$, and the sum in the pedix is to be interpreted as element-wise. See \cref{app:A} for a detailed derivation of \cref{eq:DownOut}. 
\begin{algorithm}[H]
\caption{\label{alg:Downsampling}Quantum downsampling}\vspace{2pt}
\textbf{Input} State $\ket{\Psi}_E$ \Comment{Encoding register ($n_E$ qubits)} \\[1pt]
\textbf{Parameters} \\
Integer $d$ \Comment{\# of signal dimensions }\\
Integer $\Tilde{n}< n_0 $ \Comment{Discarding parameter}
\\[1pt]
\textbf{Protocol}
\begin{algorithmic}[1]
\State apply $H^{\otimes n_E}$ 
\State apply $\mathcal{F}_E$ \vspace{1.6pt} 
\For{$0\leq s \leq d-1$}\Comment{Discarding rule}
	   \If{$n_0 - \tilde{n} \leq q \leq n_0 - 1$}
		\State{discard the $q$th qubit} 
	\EndIf
\EndFor
\State apply $\mathcal{F}^\dagger_D$ \vspace{1.6pt} 
\State apply $H^{\otimes n_D}$
\end{algorithmic}
\textbf{Result} State $\rho_D$ \Comment{Downsampled register ($n_D$ qubit)}
\end{algorithm}
As we show in \cref{fig:BlockAverage}, our protocol corresponds to a \textit{block-wise averaging operation}: it averages the input within non-overlapping hyper-cubic blocks of side $\Tilde{N}$. This is equivalent to performing a discrete convolution between $\mathcal{S}$ and a $d$-dimensional rectangular filter - whose shape arises due to the combination of the Hadamard gates and the discarding operations - with stride $\widetilde{N}$ along each axis, i.e. only retaining outputs at coordinates that are interger multiples of $\widetilde{N}$ (see \cref{app:A}). The higher the downsampling ratio of \cref{eq:CompressionRatio}, the higher the loss in output quality, as more information is averaged. Reducing the sampling rate below the Nyquist limit (twice the highest frequency component of the signal) can produce digital distortions, e.g. aliasing, \cite{art:ShannonComms}, which arises when low and high-frequency Fourier components overlap at the output. As such, our protocol is better suited to processing signals with negligible high-frequency modes in their spectra, such as traditional digital pictures, while its performances are limited for signals experiencing sharp transitions in intensities.
\begin{figure*}
    \centering
    \subfloat[]{\includegraphics[valign = c, width = 0.55\textwidth]{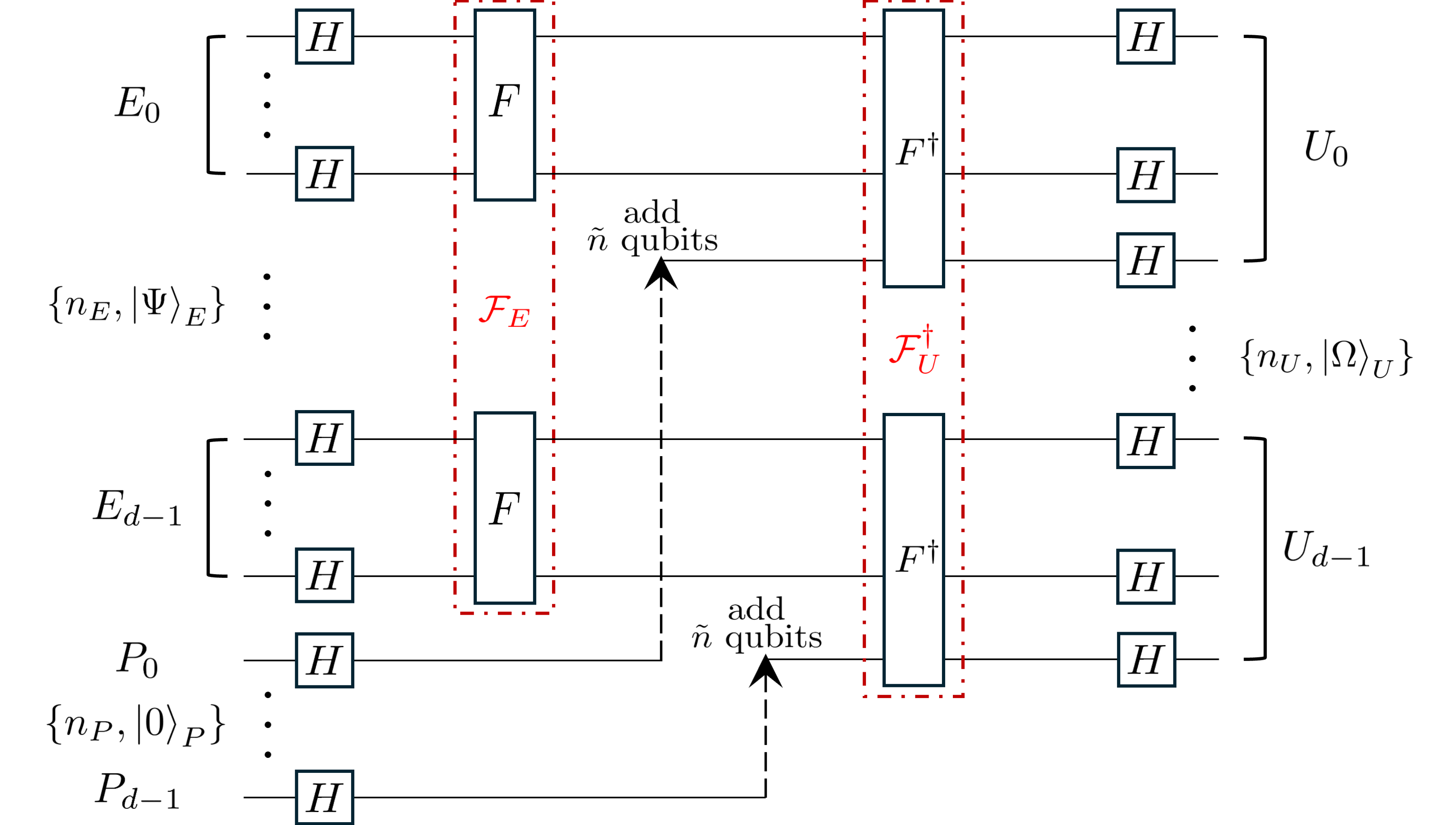}\label{fig:UpCircuit}}%
    \subfloat[]{\includegraphics[valign = c, width = 0.425\textwidth]{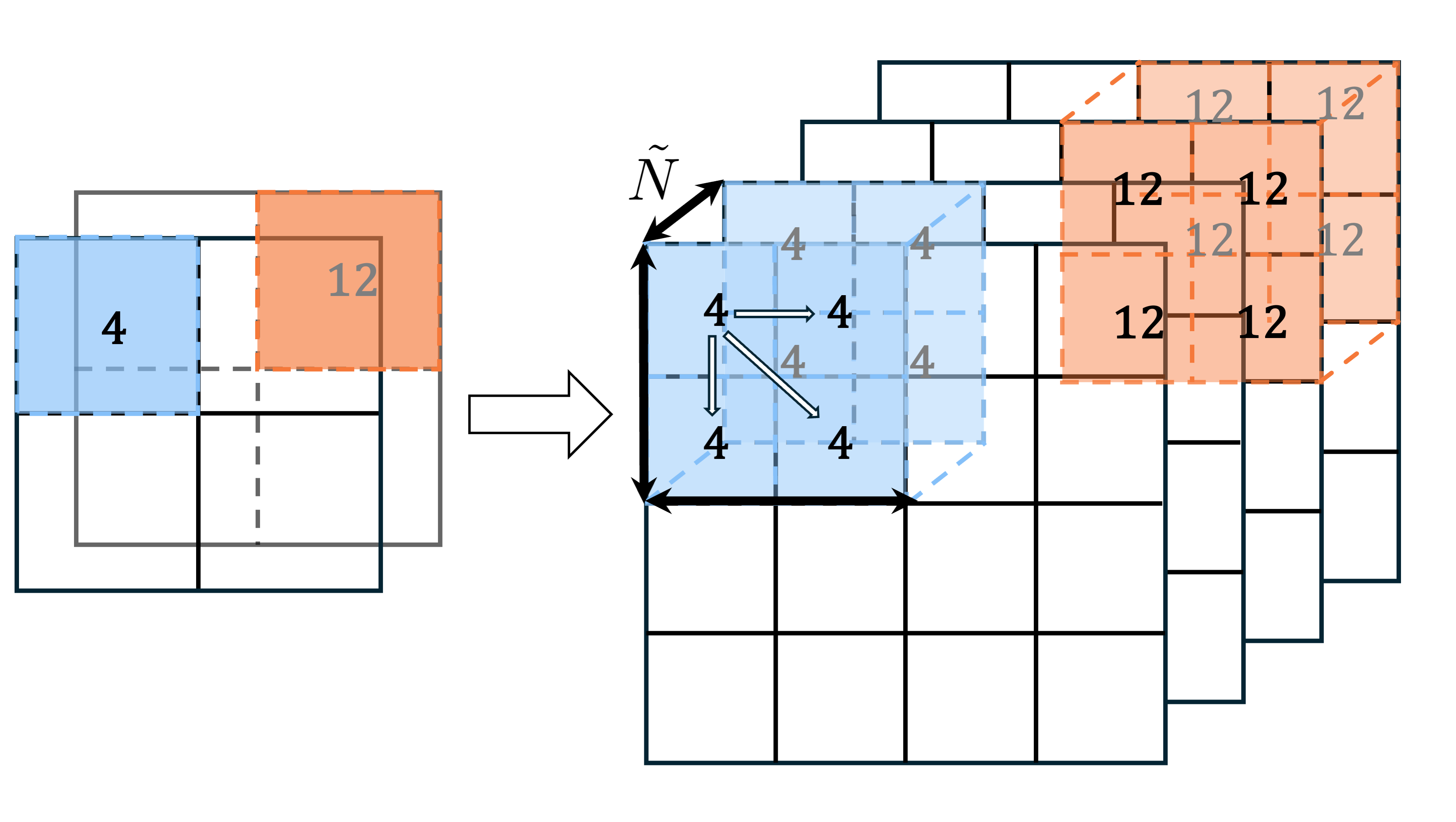}\label{fig:NearestNeighborhood}}
    
    \caption{Quantum upsampling of a $d$-dimensional signal encoded in the probabilities of register $E$, composed of $d$ subregisters $E_i$, each containing $n_0$ qubits. (a) Circuit implementation of the protocol: for the $i$-th axes, 
    $\tilde{n}$ additional qubits (the subregister $P_i$) are appended to $E_i$. The upsampled state $\ket{\Omega}_U$, encoded in $d$ subregisters $U_i$, consists of $n_0 + \tilde{n}$ qubits. In contrast to downsampling, this protocol is purity-preserving. (b) Effect on the underlying classical signal, expanded from $N_0$ to $N_0\, 2^{\tilde{n}}$ samples per axis. The protocol acts as a nearest neighborhood interpolation: if one neglects the normalization, the input values are duplicated along all axes, for a number of times determined by the padding parameter.}
\end{figure*}%

\textbf{Quantum upsampling} Complementary to the downsampling procedure, the upsampling algorithm proceeds in the reciprocal way: it expands the signal by interpolating existing samples with additional encoding resources, called \textit{padding qubits}. In signal processing, padding means expanding an input with new data of no significant content \cite{book:SmithDFT}. Similarly, our protocol calls for $n_P = d\,\tilde{n}$ qubits, i.e. the \textit{padding register} $P$, to increase the size of the encoding one. Together, these form the \textit{upsampled register} $U$, made of $n_U = d \, n_1 $ qubits, with $n_1 = n_0 + \tilde{n}$. Each padding qubit is identified by the pair $(i,p)$, where $0\leq i\leq d-1$ and $0\leq p\leq \Tilde{n}-1$. $P$ can be seen as the composition of $d$ subregisters $\{P_i\}$, one per signal axis, each expanding the corresponding encoding subregister $E_i$.
\begin{figure*}[ht]
    \centering
    \begin{tikzpicture}
       
        \node (base) at (0, 0) {\includegraphics[width=0.33\textwidth]{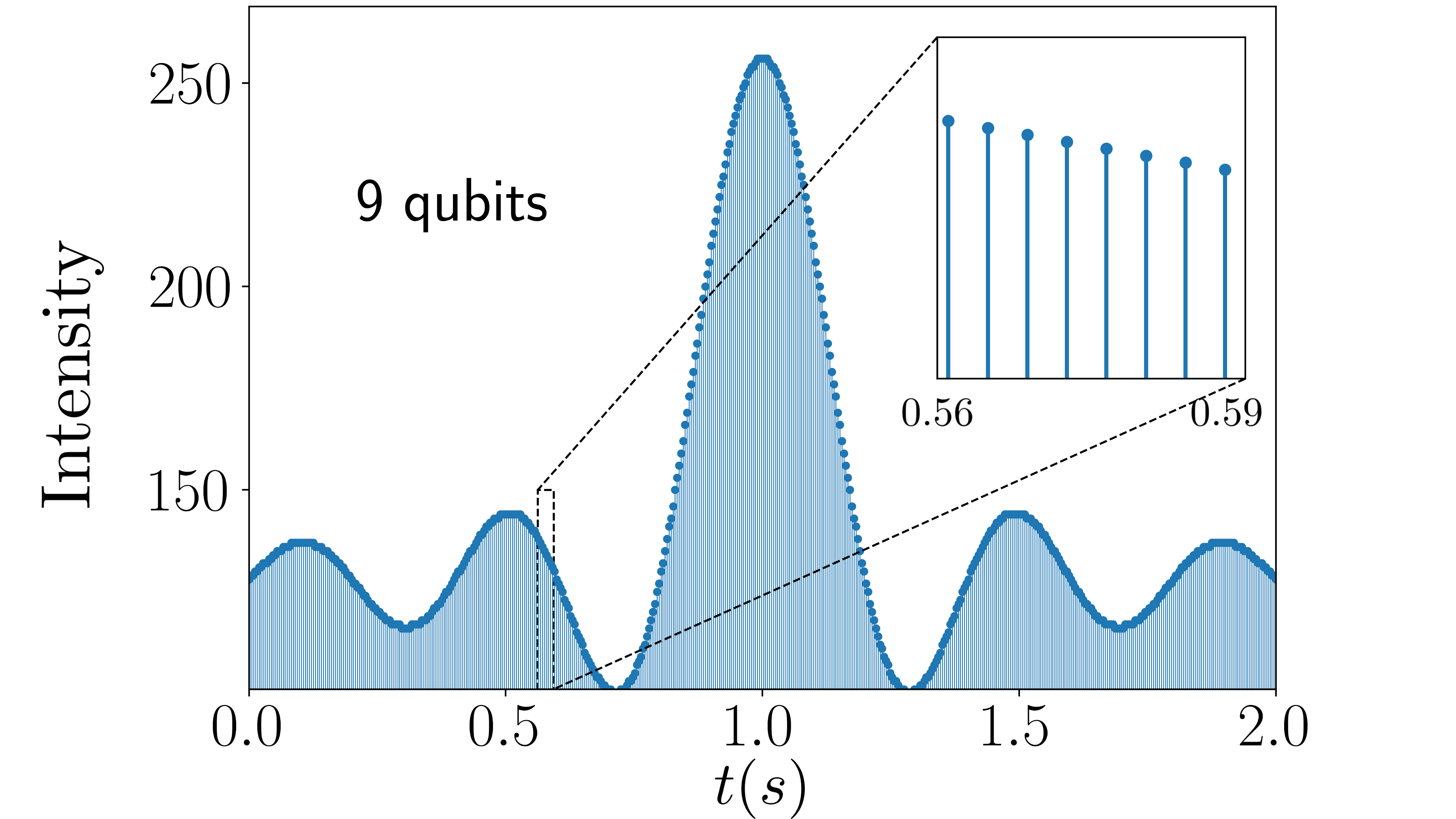}};

        \node (down) at (6, 0) {\includegraphics[width=0.33\textwidth]{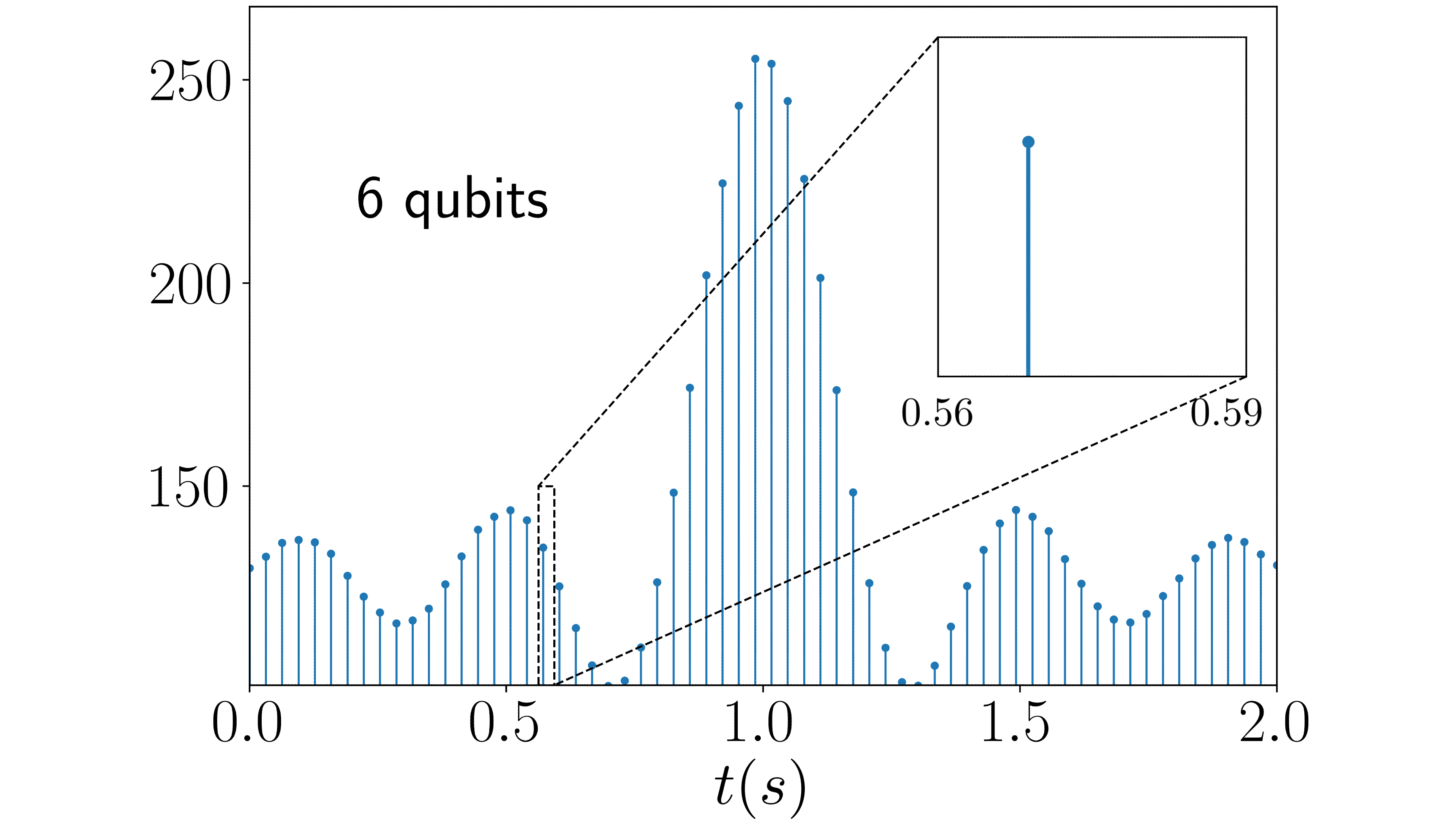}};
       
        \node (up) at (12, 0) {\includegraphics[width=0.33\textwidth]{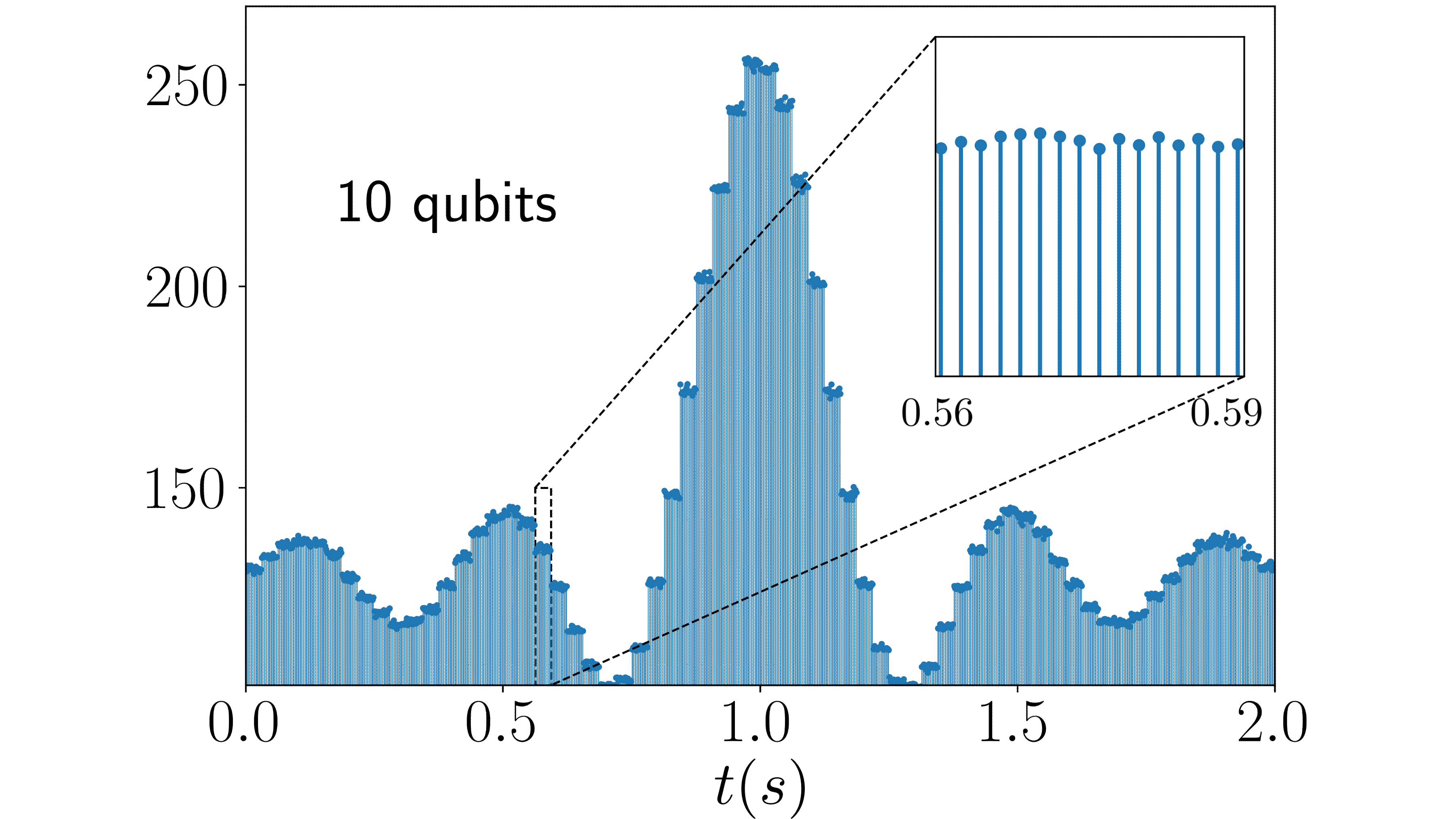}};

        \draw[<-, thick] ([xshift= 5pt] base.east) -- ([xshift= -5pt] down.west)
        node[midway, below] {\small\cref{alg:Downsampling}};

        \draw[<-, thick] ([xshift=5pt] down.east) -- ([xshift=-5pt] up.west)
        node[midway, below] {\small\cref{alg:Upsampling}};
    \end{tikzpicture}
   \caption{\label{fig:Example}Simulation of quantum resampling. A truncated sinc function (shifted by a unit value along the $y$-axis), is initially sampled over $256$ possible values at rate $\mathsf{f}_E=\SI{256}{\hertz}$ in $[0\,s,2\,s] $ and encoded in a register of $9$ qubits (left), whose computational basis indexes are obtained by discretizing and normalizing the $x$-axis (the sampling interval). The signal is first downsampled to a rate $\mathsf{f}_D=\SI{32}{\hertz}$ (6 qubits) via \cref{alg:Downsampling} (center) and then upsampled to $\mathsf{f}_U=\SI{512}{\hertz}$ (10 qubits) beyond the original resolution, employing \cref{alg:Upsampling} (right). The insets at $[0.56\, s, \,0.59\,s]$ show the effects of block-averaging and nearest neighbour interpolation. For the latter, upsampled values vary due to artefacts and statistical fluctuations. The simulation is conducted with \textit{Qiskit Aer} \cite{code:qiskit2024} and $256^2 \times 2^{n}$ shots, with $n$ being the number of output qubits.}
\end{figure*}%

The algorithm operates as follows. Initialize $P$ to $\ket{0}_{P} = \ket{0}^{\otimes n_P}$. Apply a set of Hadamard gates $H^{\otimes n_U}$ to all qubits. A MD-QFT $\mathcal{F}_E$ is taken on $E$, followed by a \textit{padding operation}: for each axis, $\tilde{n}$ qubits are moved from $P$ to the bottom of the corresponding subregister, i.e. in the most significant position. 
This operation is equivalent to introducing new zero-valued high-frequency components into the signal, expanding its spectrum. Depending on the circuit layout and its connectivity, the above padding can be actively achieved by a series of subsequent SWAP gates, appropriately shifting the qubits of the various subregisters to directly connect $E_i$ and $P_i$. Finally, perform an inverse MD-QFT $\mathcal{F}_U^\dagger$ on the full expanded register $U$, and apply a set of Hadamard gates $H^{\otimes n_U}$. The circuit representing \cref{alg:Upsampling} is depicted in \cref{fig:UpCircuit}.

The upsampled quantum state $\ket{\Omega}_U$ encodes $N_0^{d}\,2^{d\,\tilde{n}}$ entries, with each axes expanded by a factor  $2^{\tilde{n}}$ and sampling rates $\mathsf{f}_{i}^{U}=\mathsf{f}_{i}^{D}\,\widetilde{N}$. In contrast to downsampling, which is a lossy operation, the upsampling scheme is unitary, always yielding pure states at its output and thus fully preserving the amplitude encoding. Once again, output retrieval requires the knowledge of the probabilities $p_\mathbf{w}$ of observing $\ket{\mathbf{w}}_U$ at the output of the upsampled register, for all  $\mathbf{w} = (w_0,w_1, ..., w_{d-1})$ and $w_i\in\{0,N_1\}$. These are related to the input via\begin{equation}
    \label{eq:UpOut}
    p_\mathbf{w} =\frac{\mathcal{S}_{\overline{\mathbf{w}}}}{I_U} \, \, ,
\end{equation}
where $\overline{\mathbf{w}}=(w_0\,\text{mod}\,N_0,\, ...,\, w_{d-1} \,\text{mod}\,N_0,)$ , $I_U = I_E\, N_1^d $ being the output rescaled intensity and $x\,\text{mod}\,N_0$ denotes the remainder of $x$ when divided by $N_0$, so that new indices wrap around the original domain.
\begin{algorithm}[H]
\caption{\label{alg:Upsampling}Quantum upsampling}\vspace{2pt}
\textbf{Input} State $\ket{\Psi}_E$ \Comment{Encoding register ($n_E$ qubits)} \\
[1pt]
\textbf{Parameters} \\
Integer $d$ \Comment{\# of signal dimensions }\\
Integer $\tilde{n}$ \Comment{Padding/axis size}
\\[1pt]
\textbf{Protocol}
\begin{algorithmic}[1]
\State initialize padding register $P$ to $\ket{0}_P$ \Comment{ $n_P = d\,\tilde{n}$ qubits}
\State apply $H^{\otimes n_U}$
\State apply $\mathcal{F}_E$\vspace{1.6pt} 
\For{$0\leq s \leq d-1$}\Comment{Padding rule}
\For{$0\leq p \leq \tilde{n}-1$}
		\State{append $p$th qubit to the bottom of $E_i$ }
\EndFor
\EndFor
\State apply $\mathcal{F}^\dagger_U$  \vspace{1.6pt} 
\State apply $H^{\otimes n_U}$
\end{algorithmic}
\textbf{Result} State $\ket{\Omega}_U$ \Comment{Upsampled register ($n_U$ qubits)} 
\end{algorithm}
As shown in \cref{fig:NearestNeighborhood}, \cref{eq:UpOut} describes a nearest neighbor interpolation scheme, in which each data sample is repeated for $\widetilde{N}$ times along each axis. The resulting signal is piece-wise constant (within a neighbourhood of $\widetilde{N}^d$ entries) making the protocol expecially suited for sharp transitions, which are optimally preserved by this kind of interpolation technique \cite{book:Russ2018image}. An explicit derivation of this result is shown in \cref{app:B}.

Alternatively, upsampling can be implemented by substituting Steps 1 and 7 of \cref{alg:Upsampling} with a set of C-NOT gates for each subregister, each controlled on the most significant qubit of $E_i$ and targeting all the padding qubits $p\in P_i$. Although equivalent, \cref{alg:Upsampling} uses fewer entangling gates: it is thus preferable complexity wise. We refer to \cite{art:RamosQuInterpolation} for the specifics.

Combined, \cref{alg:Downsampling} and \cref{alg:Upsampling} allow to approximate multiple sampling rates, without explicitly re-digitizing the source signal. An example of their use for a one-dimensional digital signal is shown in \cref{fig:Example}. Both resampling schemes combine three key {components: the quantum Fourier transforms, the rectangular filter of the Hadamard gates and the tensor structure imposed by encoding multiple signal axes. Alternative schemes can be developed by modifying each of these elements. By substituting the QFTs, e.g. with the Haar or other wavelets transforms \cite{art:QWavelets}, it is possible to resample signals in a different domain (than the frequency one). Similarly, modifying the filter - by replacing the Hadamard gates with other unitaries - can produce outcomes different than block-averaging and nearest-neighbor interpolation, enabling higher-order polynomial interpolation or potentially addressing aliasing. Finally, different encodings can be explored, e.g. the flexible (FRQI) and the novel enhanced (NEQR) quantum representations used for image processing \cite{art:Le,art:Zhang_NEQR}.
\section{Complexity \label{sec:III}}
In this section, we compare the resource cost of the quantum resampling algorithms with that of their classical counterparts, i.e. block-averaging and nearest neighbor interpolation. First, consider a signal of $N_0^d$ samples (encoded in $n_E = d\,n_0$ qubits), downsampled to $N_0^d/2^{d\tilde{n}}$ points ($n_D = d\, n_1$ qubits, with $n_1 = n_0 - \tilde{n}$).
The gate complexity of our algorithm is dominated by the cost of performing a MD-QFT -  i.e. $d$ QFTs - on the encoding register, followed by its inverse on the downsampled one. Their total cost can be upper bounded at $\bigO(2dn_0^2)$, namely an exponential advantage over classical block-averaging in terms of number of operations only, as the latter is instead linear in the input samples i.e. $\bigO(N_0^d = 2^{dn_0})$. 
Advantage holds whenever the cost of state preparation scales at most polynomially with the number of encoding resources. This condition is met in several scenarios, exactly when the input is efficiently and classically integrable \cite{art:Grover-Rudolph,art:Soklakov_EfficentStatePrep,art:Kitaev_GaussianWavefuncs,art:RamacciottiStatePrep}, or approximately, using trainable quantum generator \cite{art:LLoyd_QGAL,art:Daimon_QuCircuit4AmplitudeEncoding,art:Dallaire-Demos_QGAN}. Similarly, preparation schemes leveraging quantum memory and encoding systems, \cite{art:Giovannetti_qRAM, art:RoncalloQJPEG} offer efficient implementations, bypassing state preparation. 

The overall complexity of our schemes must account for the cost of output recovery, i.e.: the full knowledge of the probability distribution described by \cref{eq:DownOut}.
Assuming that the input intensity $I_E$ can be stored in a classical memory during the encoding, we can estimate the output values as $O_\mathbf{m} = I_D \,\,  p_\mathbf{m}$, with $I_D$ being the downsampled intensity and $0\leq O_\mathbf{m}\leq L-1$ $\forall \mathbf{m}$. Although easily liftable, this assumption simplifies our analysis. Let $M$ denote the total number of shots, i.e. experimental repetitions of the sequence
\[
    \text{encoding} \xrightarrow{} \text{downsampling} \xrightarrow{} \text{measurement} \,\, ,
\] and let $f_\mathbf{m}$ be the observed occurence frequency for the $\mathbf{m}$-th outcome. For each outcome, the measurement statistics is effectively modelled by a Bernoulli trial : we either obtain the $\mathbf{m}$-th outcome with success probability $p_\mathbf{m}$, or not with failure probability of $1-p_\mathbf{m}$. Under the normal approximation and at $98\%$ confidence level \cite{book:Rotondi}, we can estimate the output as 
\begin{equation}
    O_\mathbf{m} = I_D \left(f_\mathbf{m} \pm 2\sqrt{\frac{f_{\mathbf{m}}(1-f_{\mathbf{m}})}{M}} \,\right)\,\, ,
\end{equation}%
with mean square error (MSE) $\Delta O^2_\mathbf{m} = 4\, I_D^2 f_{\mathbf{m}}(1-f_{\mathbf{m}})/M  $. Consider the arithmetic average of the MSE over all sample values, $ \delta^2 =2^{ -d\, n_1}\sum_{\mathbf{m}} \Delta O^2_\mathbf{m}$. Similarly, let $\langle O \rangle = 2^{ -d\, n_1}\sum_{\mathbf{m}} O_\mathbf{m}$ be the average output sample. Within our assumptions, $\langle O \rangle = 2^{ -d\, n_1} I_D = 2^{ -d\, n_0} I_E  $ , namely $\langle O \rangle$ is independent of the signal size, and thus a property of the underlying source process. 
Then, exploiting the AM-QM inequality \cite{chap:SedrakyanHMIneq} (which follows from the Cauchy-Schwartz inequality on real and positive-valued vectors) we find
\begin{equation}
    \delta^2 \leq \frac{4 \langle O 
\rangle ^2 }{M} 2^{d n_1}\,\,.
\end{equation}
whose detailed derivation can be found in \cref{app:C}. 
Thus, the number of shots required by a full output reconstruction with mean uncertainty $\delta^2 $ is 
\begin{equation}
    \label{eq:NumShots}
    M = \bigO\left(4\langle O 
\rangle ^2 \delta^{-2 }2^{d n_1}\right)\,\, .
\end{equation}
We further specialize to digital signals, for which samples can take only $L$ possible values: when these are maximal and uniformly distributed, we have $M=\bigO\left(4L ^2 \delta^{-2 }2^{d n_1}\right)$, i.e. an average of $4L^2$ shots has to be collected per sample.
The latter bound is looser than \cref{eq:NumShots}, but it provides a ``worst-case scenario" conservative estimation, valid for all outputs.
\begin{figure}[]
\centering
    \subfloat[]{\includegraphics[width=0.5\textwidth]{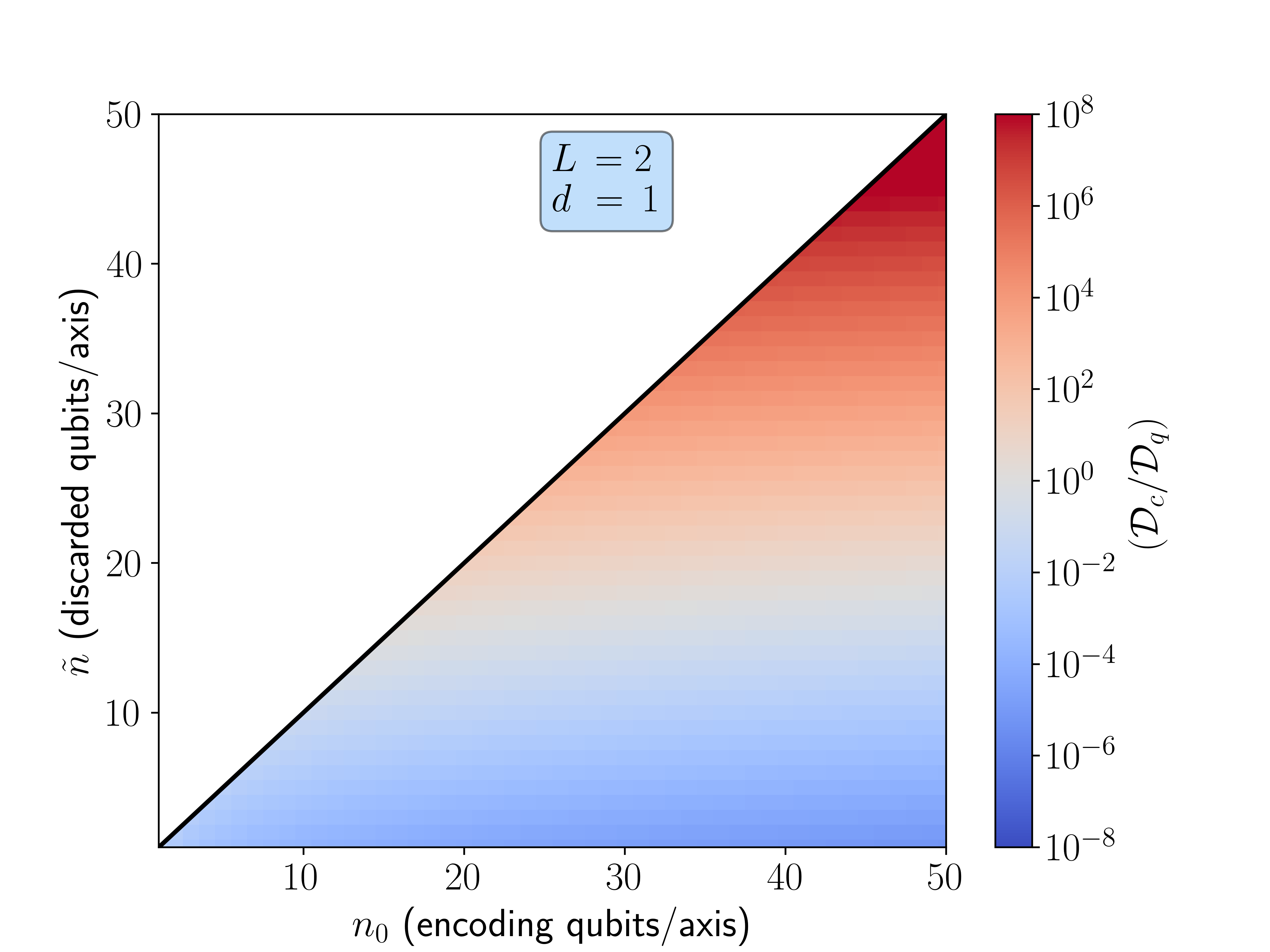}}
    \vspace*{-20pt}
    \subfloat[]{\includegraphics[width=0.5\textwidth]{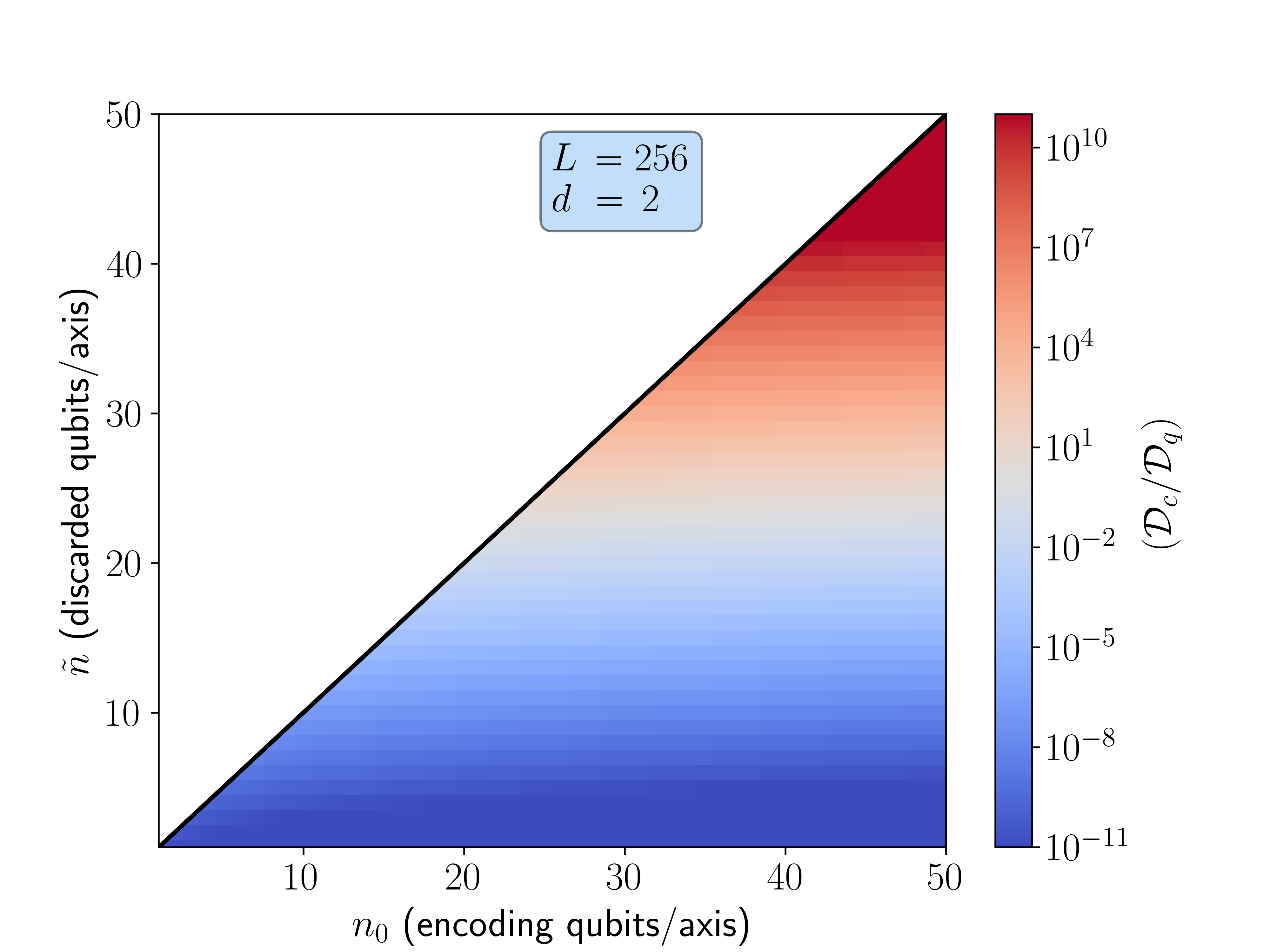}}
    
    \caption{\label{fig:Advantage}Quantum advantage bounds for the downsampling algorithms, as a function of the number of encoding and discarded qubits per signal axis. The dotted red line and the solid black one represent the lower and upper bounds of \cref{eq:BoundsForAdvantage}, respectively. The colormap expresses the ratio between classical and quantum cost $(\mathscr{D}_{\text{c}} /\mathscr{D}_{\text{q}})$, taken as figure of merit for quantifying advantage. (a) One-dimensional binary signal. (b) Two-dimensional 8-bit signal (e.g. a traditional grey-scale digital image). In both cases, the averaged MSE is set to $\delta^2 = 1/L^2$, i.e. by requiring fluctuations to be no larger than the bit-resolution.}
\end{figure}
 
Alternatively, the reconstruction can be achieved without prior knowledge of the input intensity, by normalizing the output frequencies to the maximum one, i.e. $\max_\mathbf{m} f_\mathbf{m}$. Nonetheless, the asymptotic scaling of $M$ remains comparable to \cref{eq:NumShots}, which still provides a more general bound. Both procedures can be optimized by tracking the fluctuations at the output, accumulating statistics until the desired error threshold $\delta^2$ is reached. Combining the gate and statistical costs of the algorithm (the latter taken in the worst-case scenario), we get its overall complexity, $\mathscr{D}_{\text{q}}=\bigO(8dL^2 \delta^{-2} n_0^2 \,2^{d\,n_1})$. 
An advantage holds whenever $\mathscr{D}_{\text{q}}<\mathscr{D}_{\text{c}}$, with  $\mathscr{D}_{\text{c}}=\bigO(2^{dn_0})$ being the classical cost, namely
\begin{equation}
\label{eq:BoundsForAdvantage}
\frac{1}{d}\left[2c +3+2\log_2 n_0 + \log_2 \left(\frac{d}{\delta^2}\right)\right]\leq \tilde{n} <  n_0\,\, ,
\end{equation}
where the bit depth $c = \log_2 L$ indicates the number of classical bits encoding the output value. Here, the upper bound represents the impossibility of discarding more qubits than those initially considered. As shown in \cref{fig:Advantage}, the advantage (namely the ratio between classical and quantum complexities) grows approximately with the exponential of the downsampling ratio, i.e. $\bigO(n_0^2 2^{n_E/n_D})$, and thus increases with both the input sampling rate and the number of qubits discarded. As an example, \cref{alg:Downsampling} can be employed to enhance the signal-to-noise ratios in oversampled noisy signals, as higher downsampling ratios are achievable without violating the Nyquist limit.

Analogous arguments hold for the upsampling protocol, in which the signal is expanded from $N_0^d$ ($n_E$ qubits) to $N_0^d \, 2^{d\tilde{n}}$ samples  ($n_U =  d\,n_1$ qubits, where now $n_1 = n_0 + \tilde{n}$). Similarly, \cref{alg:Upsampling} has a combined gate and statistical complexity of $\mathscr{U}_{\text{q}}=\bigO(8dL^2\delta^{-2 } n_1^2 \,2^{d\,n_1})$, where the cost of padding is dominated by the QFTs and is thus negligible. Conversely, classical nearest neighbour interpolation scales as $\mathscr{U}_{\text{c}} = \bigO(2^{d\,n_1})$. Independently of the output size, we get $\mathscr{U}_{\text{q}}>\mathscr{U}_{\text{c}}$, meaning that the quantum upsampling protocol shows no advantage per se. 

Both algorithms can work as subroutines in more complex tasks, such as image edge detection \cite{art:Yao} or classification - as it is often the case in classical signal processing scenarios - bypassing the cost of output reconstruction and recovering the full computational advantage. Additionally, our protocols can serve as pre- and post-processing layers in variational quantum models \cite{,art:Gyongyosi_Optimization,art:Gyongyosi_CircuitDepth}: inputs of varying sizes can be resampled to fit shallow, small-qubit circuits (reducing training cost), or to match larger-width devices, bypassing a costly retraining on each input size. In distributed multi-node architectures and quantum-internet settings \cite{art:Gyongyosi_ScalableDistributed,art:Gyongyosi_AdvQuInternet,art:Gyongyosi_NetQuServices}, our sample-rate conversion schemes can adjust the qubit dimension of transmitted states to match each node encoding capacity. Source nodes downsample to send a coarse-grained state, intermediate nodes aggregate partial results, and target nodes upsample to a full-width register for subsequent global operations. Similarly to \cite{art:RamosQuInterpolation}, \cref{alg:Upsampling} can enhance state preparation schemes - especially those lacking in scalability - by interpolating few (efficiently prepared) amplitudes to higher-dimensional registers. In this scenario, no reconstruction is required, and the gate overhead added by \cref{alg:Upsampling} scales quadratically in $n_1$.

The resource requirement of our algorithms can be reduced through a patch-based approach. Specifically, a $N_0^d$ signal can be split into $ \mathcal{B}= N_0/N_b $ 
smaller, non-overlapping patches of side $N_b$, which can be processed independently by \cref{alg:Downsampling,alg:Upsampling}. Such procedure is best suited to slowly varying signals, as it requires qubits to be discarded (appended) for each patch, eventually increasing aliasing.  This strategy restricts the size of the QFTs to match that of the register encoding the patch, reducing the circuit size with improved NISQ-compatibility \cite{art:Preskill_NISQ}.

\section{Conclusions}
In this work, we introduced two reciprocal quantum algorithms for multidimensional signal resampling. Combining amplitude encoding with the QFT, our protocols downsample (or upsample) all the multidimensional signal axes in parallel, modifying the number of most significant encoding qubits. We provided closed formulae and a detailed derivation of their outputs. Both downsampling and upsampling demonstrate exponential speedups over classical methods, in terms of operations only. For the former, an advantage persists despite the overhead introduced by the necessity of repeating the procedure to retrieve the full output. For the latter, the same overhead completely offsets the quantum speedup, attaining the same resource cost of its classical counterpart. Despite this limitation, the unitarity of upsampling provides easier compatibility with subsequent quantum routines, e.g. in state preparation.

Our framework focuses on frequency, but it can be adapted to different forms of quantum (or quantum-inspired) resampling, e.g. beyond the frequency domain, introducing higher-order filters (e.g. polynomial or spline interpolations) or different encoding schemes. Indeed, our results pave the way towards a comprehensive quantum library of such methods, whose range of applicability goes beyond signal processing, namely in quantum machine learning  \cite{book:Petruccione, art:Kankeu}, e.g. when compressing the parameter space in kernel methods.

\section*{Code availability}
The underlying code employed  for this study is openly available in GitHub \cite{rep:QuFRes}. 

\section*{Acknowledgments}
E.T. acknowledges support from the PNRR MUR Project PE0000023-NQSTI. S.R. acknowledges support from the PRIN MUR Project 2022RATBS4. L.M. acknowledges support from from the U.S. Department of Energy, Office of Science, National Quantum Information Science Research Centers, Superconducting Quantum Materials and Systems Center (SQMS) under Contract No. DE-AC02-07CH11359. C.M. acknowledges support from the National Research Centre for HPC, Big Data and Quantum Computing, PNRR MUR Project CN0000013-ICSC. E.T. is grateful to G.F. Conte, L. Frau, A.E. Mazzarone, D. Rinaldi and D. Tona for the helpful feedback and discussions. 

\vspace*{2cm}

\bibliography{refs.bib}

@misc{rep:QuFRes,
	author = {Emanuele Tumbiolo},
	url = {https://github.com/GitTumb/QuFRes},
	note = {\url{https://github.com/GitTumb/QuFRes}},
    year = {2025}
}

@book{book:Nielsen, 
	 place={Cambridge}, 
	 title={Quantum Computation and Quantum Information: 10th Anniversary Edition}, 
	 doi = {10.1017/CBO9780511976667}, 
	 publisher={Cambridge University Press},
	 author={Nielsen, Michael A. and Chuang, Isaac L.},
	 year={2010}
}

@book{book:Rotondi,
  	title={Probability, Statistics and Simulation: With Application Programs Written in R},
  	author={Rotondi, A. and Pedroni, P. and Pievatolo, A.},
  	doi={10.1007/978-3-031-09429-3},
  	url={https://doi.org/10.1007/978-3-031-09429-3},
  	year={2022},
  	publisher={Springer}
}

@book{book:Yan,
  	title = {Quantum Image Processing},
  	author = {Fei Yan and Salvador E. Venegas-Andraca},
  	doi = {10.1007/978-981-32-9331-1},
	url = {https://doi.org/10.1007/978-981-32-9331-1},
  	year = {2020},
  	publisher = {Springer}
}

@book{book:Petruccione,
  	title = {Machine Learning with Quantum Computers},
  	author = {Schuld, Maria and Petruccione, Francesco},
  	doi = {10.1007/978-3-030-83098-4},
	url = {https://doi.org/10.1007/978-3-030-83098-4},
  	year = {2021},
  	publisher = {Springer}
}

@misc{art:Latorre,
      title={Image compression and entanglement}, 
      author={Jose I. Latorre},
      year={2005},
      eprint={quant-ph/0510031},
      archivePrefix={arXiv},
      primaryClass={quant-ph}
}

@misc{art:Grover-Rudolph,
      title={Creating superpositions that correspond to efficiently integrable probability distributions}, 
      author={Lov Grover and Terry Rudolph},
      year={2002},
      eprint={quant-ph/0208112},
      archivePrefix={arXiv},
      primaryClass={quant-ph}
}

@article{art:Yan,
	author = {Yan, Fei and Iliyasu, Abdullah M. and Venegas-Andraca, Salvador E.},
   	title = {A survey of quantum image representations},
  	journal = {Quantum Inf. Process.},
   	year = {2016},
  	month = {Jan},
    volume = {15},
    number = {1},
    pages = {1-35},
    doi = {10.1007/s11128-015-1195-6},
    url = {https://doi.org/10.1007/s11128-015-1195-6}
}

@inproceedings{art:Venegas,
	author = {Salvador E Venegas-Andraca and Sougato Bose},
	title = {Storing, processing, and retrieving an image using quantum mechanics},
	volume = {5105},
	booktitle = {Quantum Information and Computation},
	publisher = {SPIE},
	pages = {137 -- 147},
	year = {2003},
	doi = {10.1117/12.485960},
	URL = {https://doi.org/10.1117/12.485960}
}

@article{art:Le,
	title={A flexible representation of quantum images for polynomial preparation, image compression, and processing operations},
  	author={Phuc Quang Le and Fangyan Dong and Kaoru Hirota},
  	journal={Quantum Inf. Process.},
  	year={2011},
  	volume={10},
  	pages={63-84},
  	doi={10.1007/s11128-010-0177-y},
  	url={https://doi.org/10.1007/s11128-010-0177-y}
}

@article{art:Zhang_NEQR,
  	title={{NEQR}: a novel enhanced quantum representation of digital images},
 	author={Yi Zhang and Kai Lu and Yinghui Gao and Mo Wang},
  	journal={Quantum Inf. Process.},
  	year={2013},
  	volume={12},
  	pages={2833 - 2860},
  	doi={10.1007/s11128-013-0567-z},
  	url={https://doi.org/10.1007/s11128-013-0567-z}
}

@article{art:Yao,
	title = {Quantum Image Processing and Its Application to Edge Detection: Theory and Experiment},
  	author = {Yao, Xi-Wei and Wang, Hengyan and Liao, Zeyang and Chen, Ming-Cheng and Pan, Jian and Li, Jun and Zhang, Kechao and Lin, Xingcheng and Wang, Zhehui and Luo, Zhihuang and Zheng, Wenqiang and Li, Jianzhong and Zhao, Meisheng and Peng, Xinhua and Suter, Dieter},
  	journal = {Phys. Rev. X},
  	volume = {7},
 	issue = {3},
  	pages = {031041},
  	numpages = {14},
  	year = {2017},
  	month = {Sep},
  	publisher = {American Physical Society},
  	doi = {10.1103/PhysRevX.7.031041},
  	url = {https://doi.org/10.1103/PhysRevX.7.031041}
}

@inproceedings{art:Hales,
 	author= {Hales, L. and Hallgren, S.},
  	booktitle={Proceedings 41st Annual Symposium on Foundations of Computer Science}, 
  	title={An improved quantum {F}ourier transform algorithm and applications}, 
  	year={2000},
  	volume={},
  	number={},
  	pages={515-525},
  	doi={10.1109/SFCS.2000.892139},
  	url = {https://doi.org/10.1109/SFCS.2000.892139}
}

@article{art:Zhou,
  	title={Quantum realization of the bilinear interpolation method for {NEQR}},
  	author={Ri-Gui Zhou and Wenwen Hu and Ping Fan and Hou Ian},
  	journal={Sci. Rep.},
  	year={2017},
  	volume={7},
  	pages = {2511},
  	doi = {10.1038/s41598-017-02575-6},
  	url = {https://doi.org/10.1038/s41598-017-02575-6}
}

@article{art:Preskill_NISQ,
	title = {Quantum Computing in the {NISQ} era and beyond},
	journal = {Quantum},
	year = {2018},
	volume = {2},
	pages = {79},
	author = {John Preskill},
	doi = {10.22331/q-2018-08-06-79},
	url = {https://doi.org/10.22331/q-2018-08-06-79}
}

@article{art:Giovannetti_qRAM,
  	title = {Quantum Random Access Memory},
  	author = {Giovannetti, Vittorio and Lloyd, Seth and Maccone, Lorenzo},
  	journal = {Phys. Rev. Lett.},
  	volume = {100},
  	issue = {16},
  	pages = {160501},
  	numpages = {4},
  	year = {2008},
  	month = {Apr},
  	publisher = {American Physical Society},
  	doi = {10.1103/PhysRevLett.100.160501},
  	url = {https://doi.org/10.1103/PhysRevLett.100.160501}
}

@article{art:RoncalloQJPEG,
  title={Quantum JPEG},
  author={Roncallo, Simone and Maccone, Lorenzo and Macchiavello, Chiara},
  journal={AVS Quantum Sci.},
  volume={5},
  number={4},
  year={2023},
  pages={043803},
  publisher={AIP Publishing},
  doi= {10.1116/5.0177905}
}

@article{art:ShannonComms,
  author={Shannon, C.E.},
  journal={Proc. IRE}, 
  title={Communication in the Presence of Noise}, 
  year={1949},
  volume={37},
  number={1},
  pages={10-21},
  doi={10.1109/JRPROC.1949.232969}
}

@book{book:MoirRudiments,
  title={Rudiments of Signal Processing and Systems},
  author={Moir, T.J.},
  isbn={9783030769475},
  year={2021},
  publisher={Springer International Publishing}
}

@article{art:RamosQuInterpolation,
  title = {Efficient quantum interpolation of natural data},
  author = {Ramos-Calderer, Sergi},
  journal = {Phys. Rev. A},
  volume = {106},
  issue = {6},
  pages = {062427},
  numpages = {9},
  year = {2022},
  month = {Dec},
  publisher = {American Physical Society},
  doi = {10.1103/PhysRevA.106.062427},
  url = {https://link.aps.org/doi/10.1103/PhysRevA.106.062427}
}

@book{book:SmithDFT,
  author = {Smith, Julius O.},
  isbn = {978-0-9745607-4-8},
  refid = {837679883},
  title = {Mathematics of the discrete Fourier transform (DFT)},
  year = {2007},
  publisher = {Stanford}
}

@article{art:TysonSchmidtDecomp,
   title={Operator-{S}chmidt decompositions and the Fourier transform, with applications to the operator-Schmidt numbers of unitaries},
   volume={36},
   ISSN={1361-6447},
   url={http://dx.doi.org/10.1088/0305-4470/36/24/317},
   DOI={10.1088/0305-4470/36/24/317},
   number={24},
   journal={J. Phys. A: Math. Gen.},
   publisher={IOP Publishing},
   author={Tyson, Jon},
   year={2003},
   month=jun, pages={6813–6819} }

@book{book:Russ2018image,
  title={The Image Processing Handbook},
  author={Russ, J.C. and Neal, F.B.},
  isbn={9781498740289},
  url={https://books.google.it/books?id=ROSYCgAAQBAJ},
  year={2018},
  publisher={CRC Press}

}

@book{book:LedermannComplexNums,
  title={Complex Numbers},
  author={W. Ledermann},
  isbn={9789401165709},
  doi = {10.1007/978-94-011-6570-9},
  year={2013},
  publisher={Dordrecht}
}

@inbook{chap:SedrakyanHMIneq,
author="Sedrakyan, Hayk
and Sedrakyan, Nairi",
title="The {HM-GM-AM-QM} Inequalities",
bookTitle="Algebraic Inequalities",
year="2018",
publisher="Springer International Publishing",
address="Cham",
pages="21--43",
isbn="978-3-319-77836-5",
doi="10.1007/978-3-319-77836-5_3",
url="https://doi.org/10.1007/978-3-319-77836-5_3"
}

@misc{art:Kitaev_GaussianWavefuncs,
       author = {{Kitaev}, Alexei and {Webb}, William A.},
       title={Wavefunction preparation and resampling using a quantum computer}, 
        year={2009},
        eprint={0801.0342},
        archivePrefix={arXiv},
        primaryClass={quant-ph},
}

@article{art:Soklakov_EfficentStatePrep,
  title = {Efficient state preparation for a register of quantum bits},
  author = {Soklakov, Andrei N. and Schack, R\"udiger},
  journal = {Phys. Rev. A},
  volume = {73},
  issue = {1},
  pages = {012307},
  numpages = {13},
  year = {2006},
  month = {Jan},
  publisher = {American Physical Society},
  doi = {10.1103/PhysRevA.73.012307},
  url = {https://link.aps.org/doi/10.1103/PhysRevA.73.012307}
}

@article{art:LLoyd_QGAL,
  title = {Quantum Generative Adversarial Learning},
  author = {Lloyd, Seth and Weedbrook, Christian},
  journal = {Phys. Rev. Lett.},
  volume = {121},
  issue = {4},
  pages = {040502},
  numpages = {5},
  year = {2018},
  month = {Jul},
  publisher = {American Physical Society},
  doi = {10.1103/PhysRevLett.121.040502},
  url = {https://link.aps.org/doi/10.1103/PhysRevLett.121.040502}
}

@article{art:Daimon_QuCircuit4AmplitudeEncoding,
  title = {Quantum circuit generation for amplitude encoding using a transformer decoder},
  author = {Daimon, Shunsuke and Matsushita, Yu-ichiro},
  journal = {Phys. Rev. Appl.},
  volume = {22},
  issue = {4},
  pages = {L041001},
  numpages = {6},
  year = {2024},
  month = {Oct},
  publisher = {American Physical Society},
  doi = {10.1103/PhysRevApplied.22.L041001},
  url = {https://link.aps.org/doi/10.1103/PhysRevApplied.22.L041001}
}

@article{art:Dallaire-Demos_QGAN,
  title = {Quantum generative adversarial networks},
  author = {Dallaire-Demers, Pierre-Luc and Killoran, Nathan},
  journal = {Phys. Rev. A},
  volume = {98},
  issue = {1},
  pages = {012324},
  numpages = {8},
  year = {2018},
  month = {Jul},
  publisher = {American Physical Society},
  doi = {10.1103/PhysRevA.98.012324},
  url = {https://link.aps.org/doi/10.1103/PhysRevA.98.012324}
}

@misc{code:qiskit2024,
      title={Quantum computing with {Q}iskit},
      author={Javadi-Abhari, Ali and Treinish, Matthew and Krsulich, Kevin and Wood, Christopher J. and Lishman, Jake and Gacon, Julien and Martiel, Simon and Nation, Paul D. and Bishop, Lev S. and Cross, Andrew W. and Johnson, Blake R. and Gambetta, Jay M.},
      year={2024},
      eprint={2405.08810},
      archivePrefix={arXiv},
      primaryClass={quant-ph}
}

@InProceedings{art:QWavelets,
    author="Fijany, Amirand Williams, Colin P.",
    editor="Williams, Colin P.",
    title="Quantum Wavelet Transforms: Fast Algorithms and Complete Circuits",
    booktitle="Quantum Computing and Quantum Communications",
    year="1999",
    publisher="Springer",
    pages="10--33",
}

@book{book:Oppenheim2011discrete,
  title={Discrete-Time Signal {P}rocessing},
  author={Oppenheim, A.V. and Schafer, R.W.},
  isbn={9780133002287},
  url={https://books.google.it/books?id=EaMuAAAAQBAJ},
  year={2011},
  publisher={Pearson Education}
}

@misc{art:Peng,
      title={Hybrid Quantum Downsampling Networks}, 
      author={Yifeng Peng and Xinyi Li and Zhiding Liang and Ying Wang},
      year={2024},
      eprint={2405.16375},
      archivePrefix={arXiv},
      primaryClass={quant-ph},
}

@misc{art:Stefanski,
      title={Quantum Amplitude Interpolation}, 
      author={Charlee Stefanski and Vanio Markov and Constantin Gonciulea},
      year={2022},
      eprint={2203.08758},
      archivePrefix={arXiv},
      primaryClass={quant-ph},
}

@misc{art:Kankeu,
      title = {Quantum-inspired Embeddings Projection and Similarity Metrics for Representation Learning}, 
      author ={Ivan Kankeu and Stefan Gerd Fritsch and Gunnar Schönhoff and Elie Mounzer and Paul Lukowicz and Maximilian Kiefer-Emmanouilidis},
      year={2025},
      eprint={2501.04591},
      archivePrefix={arXiv},
      primaryClass={cs.CL},
}

@article{art:RamacciottiStatePrep,
  title = {Simple quantum algorithm to efficiently prepare sparse states},
  author = {Ramacciotti, Debora and Lefterovici, Andreea I. and Rotundo, Antonio F.},
  journal = {Phys. Rev. A},
  volume = {110},
  issue = {3},
  pages = {032609},
  numpages = {10},
  year = {2024},
  month = {Sep},
  publisher = {American Physical Society},
  doi = {10.1103/PhysRevA.110.032609},
  url = {https://link.aps.org/doi/10.1103/PhysRevA.110.032609}
}

@article{art:Gyongyosi_Optimization,
  title={Training Optimization for Gate-Model Quantum Neural Networks},
  author={Gyongyosi, Laszlo and Imre, Sandor},
  journal={Scientific Reports},
  volume={9},
  number={1},
  pages={12679},
  year={2019},
  publisher={Nature Publishing Group},
  doi={10.1038/s41598-019-48892-w},
  url={https://www.nature.com/articles/s41598-019-48892-w}
}

@article{art:Gyongyosi_CircuitDepth,
  title={Circuit Depth Reduction for Gate-Model Quantum Computers},
  author={Gyongyosi, Laszlo and Imre, Sandor},
  journal={Scientific Reports},
  volume={10},
  number={1},
  pages={11229},
  year={2020},
  publisher={Nature Publishing Group},
  doi={10.1038/s41598-020-67014-5},
  url={https://www.nature.com/articles/s41598-020-67014-5}
}

@article{art:Gyongyosi_ScalableDistributed,
  title={Scalable distributed gate-model quantum computers},
  author={Gyongyosi, Laszlo and Imre, Sandor},
  journal={Scientific Reports},
  volume={11},
  number={1},
  pages={5172},
  year={2021},
  publisher={Nature Publishing Group},
  doi={10.1038/s41598-020-76728-5},
  url={https://www.nature.com/articles/s41598-020-76728-5}
}

@article{art:Gyongyosi_NetQuServices,
    author = {Gyongyosi, Laszlo and Imre, Sandor},
    doi = {10.2478/qic-2025-0006},
    url = {https://doi.org/10.2478/qic-2025-0006},
    title = {Networked Quantum Services},
    journal = {Quantum Information \& Computation},
    number = {2},
    volume = {25},
    year = {2025},
    pages = {97--140}
}

@article{art:Gyongyosi_AdvQuInternet,
  title        = {Advances in the Quantum Internet},
  author       = {Gyongyosi, Laszlo and Imre, Sandor},
  journal      = {Communications of the {ACM}},
  volume       = {65},
  number       = {8},
  pages        = {52--63},
  year         = {2022},
  doi          = {10.1145/3524455},
  url          = {https://doi.org/10.1145/3524455}
}

@book{book:Patterson_LittleEndian,
  author    = {David A. Patterson and John L. Hennessy},
  title     = {Computer Organization and Design: The Hardware/Software Interface},
  edition   = {5th},
  publisher = {Morgan Kaufmann},
  year      = {2013}
}

\appendix
\onecolumngrid
\newpage
\section{REGISTERS IN THE FREQUENCY DOMAIN}\label{app:0} In this section, we give an intuitive explanation of how our resampling algorithms operate in the frequency domain. For an amplitude-encoded quantum state, $n$-qubits support up to $2^n$ frequency bins,  with the highest bin index being $k=2^n-1$ (corresponding to ``normalized" frequency of $(2^n-1)/2^n$ cycles/sample). The upper end of the spectrum is controlled by the most significant qubits in the register. To see this,  consider a generic computational basis state $\ket{x}$ under the Quantum Fourier Transform:
\begin{equation*}
\label{eq:qft-definition}
\ket{x} \xrightarrow[QFT]{} \frac{1}{\sqrt{2^n}}
\sum_{y=0}^{2^n-1}
e^{2\pi i\,x y / 2^n}\,\ket{y}
= \bigotimes_{j=1}^{n}
\frac{\ket{0} + e^{2\pi i\,x / 2^j}\,\ket{1}}{\sqrt{2}}.
\end{equation*}
Here, each qubit $j$ carries a phase factor $e^{2\pi i\,x / 2^j}$, corresponding to a (normalized) frequency of $1/2^j$ cycles/sample. In particular:
\begin{itemize}
  \item Qubit $j=1$, namely the most significant one, carries phase $e^{2\pi i\,x/2}$, i.e. flips with a frequency $\frac{1}{2}$ cycles/sample, the highest supported by the $n$-qubit register.
  \item Qubit $j=2$, the second most significant, carries phase $e^{2\pi i\,x/2^2}$, oscillating every two samples-frequency $\frac{1}{4}$.
  \item Qubit $j=n$, the least significant qubit in the register, carries phase $e^{2\pi i\,x/2^n}$ oscillating with frequency $\frac{1}{2^n}$ cycles/sample.
\end{itemize}
In this sense, controlling the most significant qubits directly shapes the high-frequency content.  Discarding the most significant qubits removes all components with $k\ge2^{n-1}$ - namely the top half of the spectrum - averaging out the state over the fastest modes.  Conversely, appending fresh most significant qubit increases the number of bins - by addding new bits of spectral resolution - with the new high-frequency slots being initalized to zero.
\section{QUANTUM DOWNSAMPLING \label{app:A}}
In this section, we compute the step-by-step evolution of a signal downsampled via \cref{alg:Downsampling}. For simplicity, we consider $d=1$, namely a one-dimensional discrete signal $\mathcal{S} = (\mathcal{S}_0,\mathcal{S}_1,...,\mathcal{S}_{N_0-1})$, with $N_0$ samples that are encoded into a quantum register of $n_E = n_0+\log_2N_0$ qubits as 
\begin{equation}
    \label{eq:InputState}
    \ket{\Psi}_E= \sum_{e=0}^{N_0-1} \alpha_e \ket{e}\; ,
\end{equation}
where $\alpha_e = \sqrt{\mathcal{S}_e/I_E}$, $I_E = \sum_{e=0}^{N_0-1} \mathcal{S}_e$ and with $e$ labelling both classical sample values and computational basis states.
Starting from $\ket{\psi}_E$, our algorithm discards its most $\Tilde{n}$ most significant qubits, associated with the highest spatial frequencies, reducing the number of samples in the signal to $N_1 = N_0/\widetilde{N}$, where $\widetilde{N} = 2^{\Tilde{n}}, N_1 = 2^{n_1}$ and $n_1 = n - \Tilde{n} $. We use pedix $\bullet_1$ and diacritic $\Tilde{\bullet}$ to label indices running from $0$ to $N_1$ and from $0$ to $\widetilde{N}$, respectively. Conversely, any unmarked index is intended to assume values from $0$ to $N_0$, unless otherwise stated. Since for one-dimensional signals  $n_E (n_D) = n_0 (n_1)$, we suppress the $\ket{\bullet}_E$ notation.

 The quantum downsampling protocol begins by applying $n_0$ Hadamard gates 
\begin{gather}
    H^{\otimes n_0} \ket{\psi}  =  \sum_z \beta_z \ket{z}\;,
    \label{eq:BetaCoefficients} \\
    \text{where} \ \beta_z = \frac{1}{\sqrt{N_0}}\sum_e \alpha_e (-1)^{e\odot z}\; ,
\end{gather}
and with $\odot$ being the modulo-2 bitwise inner product \cite{book:Nielsen}. This is followed by a quantum Fourier transform $F_0$, performed on the full register
\begin{equation}
\label{eq:QFTStep}
    F_0\left(\sum_z \beta_z \ket{z}\right) = \frac{1}{\sqrt{N_0}}\sum_{z,k}\beta_z \omega_{N_0}^{k z} \ket{k} \; ,
\end{equation}
where $\omega_N = \exp{2 \pi i/N }$ denotes the N-th root of the imaginary unity.
Then, the $\Tilde{n}$ most significant qubits of the register are discarded, i.e. by a partial trace on the subspace $\widetilde{\mathcal{H}}$ spanned by such qubits. For simplicity, we postpone this computation, which commutes with any local operation applied to the remaining qubits. We decompose the two subregisters as 
\begin{equation}
\label{eq:k_Decomposition}
    k = N_1 \Tilde{k} + k_1 \; ,
\end{equation}
yielding $\omega_{N_0}^{kz} = \omega_{\widetilde{N}}^{\Tilde{k}z} \cdot \omega_{N_0}^{k_1 z}$, and 
\begin{equation}
    \frac{1}{\sqrt{N_0}}\sum_{z}\underbrace{\sum_{\Tilde{k}} \omega_{\widetilde{N}}^{\Tilde{k}z} \ket{\Tilde{k}}}_{\ket{\Tilde{\chi}_z}}\otimes\underbrace{\beta_z\sum_{k_1}\omega_{N_0}^{k_1 z} \ket{k_1}}_{\ket{\chi_z^1}}\; .
\end{equation}
Alternatively, the same formula can be recovered from the Schmidt decomposition of the QFT operator \cite{art:TysonSchmidtDecomp}. Following \cref{alg:Downsampling}, we apply the inverse QFT $F_1^\dagger$ to the $n_1$-register, yielding
\begin{equation}
    \left(\mathds{1}_{\Tilde{n}}\otimes F_1^{\dagger}\right) \left(\frac{1}{\sqrt{N_0}}\sum_{z}\ket{\widetilde{\chi}_z}\otimes\ket{\chi_z^1}\right) =
 \frac{1}{\sqrt{N_0N_1}}\sum_{z}\ket{\widetilde{\chi}_z}\otimes \beta_z\sum_{k_1,t_1}\underbrace{\omega_{N_0}^{k_1 z}\omega_{N_1}^{-k_1 t_1}}_{\omega_{N_0}^{-k_1 (z - \widetilde{N} t_1)}} \ket{t_1}\; ,
\end{equation}
where $\mathds{1}_{\Tilde{n}}$ is the identity operator on the $\widetilde{\mathcal{H}}$ subspace.
The protocol ends with a final set of Hadamard gates on the $n_1$-register, getting
\begin{gather}
     \ket{\Phi} = \frac{1}{\sqrt{N_0}N_1}\sum_{z}\ket{\widetilde{\chi}_z} \otimes\ket{\lambda_z^1} \;, \\
    \text{where} \ \ket{\lambda_z} = \sum_{k_1,t_1,l_1}\beta_z\omega_{N_0}^{k_1 (z - \widetilde{N}t_1)} (-1)^{t_1\odot l_1}\ket{l_1}\; .
\end{gather}
We now explicitly discuss the discarding operation. The density operator associated with the state $\ket{\Phi}$ is 
\begin{equation}
\projector{\Phi}{\Phi}{}=\frac{1}{N_0 N_1^2}\sum_{z,z'} \projector{\widetilde{\chi}_z}{\widetilde{\chi}_{z'}}{} \otimes \projector{\lambda_z^{1}}{\lambda_{z'}^1}{} \;, 
\end{equation}
Tracing on $\widetilde{\mathcal{H}}$ gives
\begin{equation}
\label{eq:ReducedRho}
         \rho_1 = \text{Tr}_{\Tilde{\mathcal{H}}}\left[ \projector{\Phi}{\Phi}{}\right] =
         \sum_{z,z'} \sum_{\Tilde{k}}\omega_{\Tilde{N}}^{\Tilde{k}(z-z')}\projector{\lambda_z^{1}}{\lambda_{z'}^{1}}{}\; \ . 
\end{equation}
Observe that
\begin{equation}
\label{eq:DeltaZZ'}
    \sum_{\Tilde{k}}\omega_{\widetilde{N}}^{\Tilde{k}(z-z')} =
    \begin{cases}
     \widetilde{N}\quad &\text{if } z \equiv z' \mod \widetilde{N} \\
         0\quad &\text{otherwise} 
\end{cases}\; ,
\end{equation}
which acts as a modular Kronecker delta between $z$ and $z'$ \cite{book:LedermannComplexNums}.
Condition \ref{eq:DeltaZZ'} implies that we can decompose $z,z'$ into $z = \widetilde{N}j_1+\Tilde{r}$ and $z' = \widetilde{N}j^{'}_1+\Tilde{r}$, with $j_1 ( j_1')$ being the results of the division of $z$ ($z'$) by $\widetilde{N}$ and $\Tilde{r}$ the common remainder. We thus decompose the summations in $z,z'$ as $\sum_z \sum_{z'} \xrightarrow[]{} \sum_{j_1} \sum_{j_1'} \sum_{\Tilde{r}}$.
Recalling \cref{eq:BetaCoefficients}, together with the definition of the modulo-2 bitwise inner product, gives
\begin{equation}
\begin{split}
    \beta_z=\beta_{\widetilde{N}j_1+\Tilde{r}} =\sum_{e_1,\Tilde{e}}\frac{1}{\sqrt{N_0}}\alpha_{e_1,\Tilde{e}} (-1)^{j_1\odot e_1} (-1)^{\Tilde{r}\odot \Tilde{e}}\, ,
\end{split}
\end{equation}
and similarly for $\beta_{z'}$. Here, $e_1, \Tilde{e}$ and $e$ undergo $e = \widetilde{N}e_1+\Tilde{e}$, which parametrizes the integers formed of the $n_1$ most and $\Tilde{n}$ least significant bits of $e$, respectively. Then
\begin{equation}
\label{eq:ExplicitInComput}
\begin{split}
    \rho_1 = &\frac{1}{N_1^3N_0}\sum_{\mathclap{\substack{k_1,e_1\\ k_1',e_1'}}}\sum_{\Tilde{e},\Tilde{e}',\Tilde{r}}\alpha_{e_1,\Tilde{e}}\,\alpha^*_{e_1',\Tilde{e}'}(-1)^{\Tilde{r}\odot(\Tilde{e}+\Tilde{e}')}
    \omega_{N_0}^{\Tilde{r}(k_1-k_1')}\\
    &\cross \sum_{\mathclap{\substack{j_1,t_1\\ j_1',t_1'}}}
    \phi(k_1,e_1,j_1,t_1)\phi^{*}(k_1',e_1',j_1',t_1')\sum_{l_1,l_1'}
    (-1)^{t_1\odot l_1}(-1)^{t_1'\odot l_1'}\projector{l_1}{l'_1}{} \, ,
\end{split} 
\end{equation} where
\begin{equation}
    \phi(k_1,e_1,j_1,t_1) = \omega_{N_1}^{k_1(j_1-t_1)}(-1)^{(e_1 \odot j_1)}\; .
\end{equation}
Observe that $\phi$ is periodic with period $N_1$ in each of its arguments. Hence, it is invariant under the transformation 
\begin{gather}
    \begin{cases}
        j_1 \xrightarrow{} t_1 + p_1 \\ 
        j_1' \xrightarrow{}t_1' + p_1'
    \end{cases} \Rightarrow \ \ \sum_{t_1=0}^{N_1-1}\sum_{p_1=-t_1}^{N_1-1-t_1} = \sum_{p_1=0}^{N_1-1}\sum_{t_1=0}^{N_1-1}\;,
\end{gather}
and similarly for $t_1'$ and $p_1'$.
We combine this change of variable with the identity
\begin{equation}
\label{eq:HadamardDelta}
    \sum_a^{N-1} (-1)^{a\odot(c+b)} = N\delta_{c,b}\; ,
\end{equation}
where $\delta_{c,b}$ is the usual Kronecker delta - which follows from the involutivity of the Hadamard operator. Then  \cref{eq:ExplicitInComput} simplifies to
\begin{equation}
\begin{split}
    &\rho_1 = \frac{1}{\widetilde{N}N_1^2}\sum_{\mathclap{\substack{k_1,e_1\\ k_1',e_1'}}}\sum_{\Tilde{e},\Tilde{e}',\Tilde{r}} \alpha_{e_1,\Tilde{e}}\,\alpha^{*}_{e_1',\Tilde{e}'}(-1)^{\Tilde{r}\odot(\Tilde{e}+\Tilde{e}')}\omega_{N_0}^{\Tilde{r}(k_1-k_1')} \sum_{p_1,p^{'}_1}\omega_{N_1}^{k_1 p_1}(-1)^{p_1\odot e_1}\omega_{N_1}^{k_1' p_1'}(-1)^{p_1'\odot e_1^{'}}\projector{e_1}{e_1'}{} \ . 
\end{split}
\end{equation}\\
We recover the downsampled signal from the output probabilities $p(m_1) = \Tr\left[\rho_1 \projector{m_1}{m_1}{}\right]$, which in our case read
\begin{equation}
\label{eq:OutProbs_AppA}
\begin{split}
    p(m_1) = 
    \frac{1}{\widetilde{N}N_1^2}\sum_{k_1,k_1'}\sum_{\Tilde{e},\Tilde{e}',\Tilde{r}} \alpha_{m_1,\Tilde{e}}\,\alpha^{*}_{m_1,\Tilde{e}'}(-1)^{\Tilde{r}\odot(\Tilde{e}+\Tilde{e}')}\omega_{N_0}^{\Tilde{r}(k_1-k_1')} \sum_{p_1,p^{'}_1}\omega_{N_1}^{k_1 p_1}\omega_{N_1}^{k_1' p_1'}(-1)^{m_1\odot(p_1+p_1') }\; .
\end{split}
\end{equation}
Exploiting the periodicity of the sums on $p_1$ and $p_1'$ appearing, we set $p_1  \xrightarrow{} p'_1 + h_1$, getting

\begin{equation}
    \begin{cases}
        {\omega_{N_1}^{k_1 p_1}\omega_{N_1}^{k_1' p_1'} \xrightarrow{} \omega_{N_1}^{p_1' (k_1-k_1')}\omega_{N_1}^{h_1 k_1}}\vspace{0.2cm}\\ 
        
        {(-1)^{m_1\odot(p_1+p_1')}  \xrightarrow{}(-1)^{m_1\odot h_1}} 
    \end{cases} \ .
\end{equation}
We complete the computations by observing that: (i) the summations on $p_1'$ and $k_1$ give $\delta_{k_1,k_1'}$ and $\delta_{h,0}$, respectively; (2) due to \cref{eq:HadamardDelta}, the summation on $\Tilde{r}$ yields $\delta_{\Tilde{e},\Tilde{e}'}$ ; ending up with
\begin{equation}
    p(m_1) = \widetilde{N}\sum_{\Tilde{e}}\abs{\alpha_{m_1,\Tilde{e}}}^2=\sum_{\Tilde{e}} \frac{\mathcal{S}_{\widetilde{N}m_1+\Tilde{e}}}{I_D} \,\, ,
\end{equation}
where $I_D = \widetilde{N}/I_E$ .
As discussed in the main text, observe that this output can be seen as a convolution of  stride $\widetilde{N}$ between the input signal and a uniform rectangular  filter $w$ of side $\widetilde{N}$, namely
\begin{gather}
   p(m_1)=
    (\mathcal{S} * w)\bigl[\widetilde{N}m_1 \bigr]\,, \\
    \text{where}\,\,\, (\mathcal{S} * w)\bigl[j\bigr] = \sum_{k=0}^{\widetilde{N}-1} \mathcal{S}\bigl[j\bigr]\,w\bigl[j-k\bigr]\; \,\,\text{and}\,\,\,
    w[j] =  \begin{cases}
      1/I_D, & 0 \,\le\, j \,<\, \widetilde{N},\\
      0, & \text{otherwise}.
    \end{cases}
\end{gather}%
Since \cref{alg:Downsampling} processes distinct subregisters in parallel, the extension to the multidimensional case is immediate. This shows that the protocol performs a block-average of the input samples, whose output size depends on the number of qubits discarded from the original register.

\section{QUANTUM UPSAMPLING\label{app:B}}
In this section, we compute the evolution of a signal extended to a bigger support via our quantum upsampling algorithm. Without loss of generality, we consider once again the one-dimensional case, and employ the same conventions discussed in the previous section. 
The upsampling scheme is a padding of the previously considered state $\ket{\Psi}_E$  in the frequency domain, using a set of $\Tilde{n}$ uninitialized qubits, which are added to the register as the new most significant ones. This operation increases the number of qubits from $n_0$ to $n_1=n_0+\Tilde{n}$. We refer to the $\Tilde{n}$-register as padding register. Therefore, we consider as input $\ket{0}^{\otimes\Tilde{n}}\otimes\ket{\Psi}_E$.
First, the algorithm applies a set of Hadamard gates to all the qubits, producing
\begin{equation}
    \frac{1}{\sqrt{\widetilde{N}}}\sum_{\Tilde{p}} \ket{\Tilde{p}}\otimes \sum_z\beta_z \ket{z}\; ,
\end{equation}
where $\beta_z$ is defined as in \cref{eq:BetaCoefficients}. Then, we perform a QFT on the initial $n_0$ qubits, while the padding register remains unaltered. Then
\begin{equation}
    \frac{1}{\sqrt{\Tilde{N}  N_0}}\sum_{\Tilde{p}}\ket{\Tilde{p}}\otimes\sum_{z,k}\beta_z \omega_{N_0}^{kz}\ket{k}\; . 
\end{equation}
The tensor product states $\ket{\Tilde{p}}\otimes\ket{k}$ are in a one-to-one correspondence with $\ket{t_1}=\ket{N_0\Tilde{p}+k}$. This change of summation variables gives
\begin{equation}
    \frac{1}{\sqrt{\widetilde{N} N_0}}\sum_{z}\sum_{t_1}\beta_z\omega_{N_0}^{t_1 z}\ket{t_1}\; .
\end{equation}
The last two steps of the protocol consist of a full inverse QFT, followed by Hadamard gates applied to all qubits
\begin{equation}
\label{eq:PenultimStep}
\begin{split}
 \frac{1}{\sqrt{N_1^3}}\sum_z\beta_z
 \sum_{v_1,l_1}\underbrace{\sum_{t_1}\omega_{N_1}^{t_1(\Tilde{N}z-v_1)}}_{N_1\delta_{\widetilde{N}z,v_1}} (-1)^{v_1\odot l_1}\ket{l_1}\; .
\end{split}
\end{equation}
Inserting the definition of $\beta_z$ in \cref{eq:PenultimStep}, and using the same arguments that lead us to \cref{eq:BetaCoefficients}, we finally get
\begin{equation}
    \ket{\Omega}=\frac{1}{\sqrt{N_1}N_0}\sum_{e,l}\sum_{\Tilde{l}}\alpha_e \underbrace{\sum_z (-1)^{z\odot(e+l)}}_{N_0\delta_{e,l}} \ket{l}\ket{\Tilde{l}} = \frac{1}{\sqrt{N_1}}\sum_{e}\sum_{\Tilde{l}}
    \alpha_e \ket{e}\ket{\Tilde{l}}\;,
\end{equation}
where $l_1 = \widetilde{N}l+\Tilde{l}$. 
Once again, the output signal is recovered from the probabilities of each outcome in the computational basis. From the Born rule
\begin{equation}
     p(w_1) = p(\widetilde{N}w+\widetilde{w}) =\abs{(\bra{w}\otimes\bra{\widetilde{w}})\ket{\Phi}_U}^2=\frac{1}{N_1}\abs{\alpha_w}^2 = \mathcal{S}_w/I_U \ .
\end{equation}
Here $I_U = I_E \, N_1$.
This holds for any $w'_1$ such that $w'_1 \equiv w_1 \mod{\widetilde{N}}$, meaning that all the samples with indexes in the same congruence class modulo $\widetilde{N}$ will be identical. The generalization to the multidimensional case is  guaranteed by the MD-QFT acting independently on distinct subregisters, leading to \cref{eq:UpOut}. This shows that the protocol actually performs a nearest-neighbor interpolation, whose size depends on the number of padding qubits added to the initial register.

\section{STATISTICS OF OUTPUT RECONSTRUCTION\label{app:C}} 
In this section, we discuss the size of the statistical sample, i.e. the number of shots $M$ required to reconstruct the output at a given uncertainty. As stated in the main text, this cost dampens the advantage of our algorithms with respect to their counterparts, imposing a trade-off between efficiency and output quality. Conversely, the theoretical advantage is recovered whenever a full output reconstruction is not needed.

Without loss of generality, we limit our discussion to the downsampling of a one-dimensional signal, from $N_0$ ($n_0$ qubits) to $N_1= N_0/2^{\tilde{n}}$ ($n_1$ qubits), with output intensity $I_D$.
Let $p_m$ be the probability mass function associated to a computational basis measurement on the output state, with $m \in \{0,N_1 -1 \}$. Due to the probabilistic nature of our encoding, output sample values can be expressed as $O_m = I_D p_m$. Following the main text, their maximum likelihood estimator at a $98\%$ confidence level is
\begin{equation}
    O_m = I_D f_m \pm 2 I_D \sqrt{\frac{f_m (1-f_m)}{M}}
\end{equation}
where $f_m$ is again the occurrence frequency for the $m$-th output. Let $\delta^2$ be the arithmetic average of the squared standard error
over all samples, namely  
\begin{equation}
    \delta^2 = \frac{1}{N}\sum^{N_1-1}_{m=0} \Delta O_m^2 = \frac{4 I_D^2}{M}\sum^{N_1-1}_{m=0} (f_m-f_m^2)\,\, .
\end{equation}
From the AM-QM inequality \cite{chap:SedrakyanHMIneq} - i.e. applying Cauchy-Schwartz to the vector of frequencies $f_m$ and a vector of ones $(1,1,\dots,1) $  - 
\begin{equation}
    \sum_{m=0}^{N_1-1} \frac{f_m^2}{N_1} \geq \frac{1}{N^2_1}\bigg(\sum_m f_m\bigg)^2 = \frac{1}{N^2_1}\,\, ,
\end{equation}
leading us to 
\begin{equation}
\delta^2 \leq \frac{4I_D^2}{M}\left(\frac{N_1-1}{N_1^2}\right)\leq   \frac{4I_D^2}{M}\frac{1}{N_1} \ , 
\end{equation}
where the last inequality holds for sufficiently large signals, i.e. $N_1\gg1 $. Let $\langle O \rangle = I_D/N_1  $ be the average sample value in the output, then
\begin{equation}
    \label{eq:SignalDependent}
    M\leq 4 \langle O\rangle^2 N_1\delta^{-2}\,\, .
\end{equation}
\cref{eq:SignalDependent} gives a signal-dependent bound. Conversely a looser, all-purpose bound directly follows when the output samples have the highest possible values, namely $\langle O \rangle = L-1$. 

\end{document}